\documentclass[apsrev,prb,twocolumn,superscriptaddress]{revtex4-2}
\usepackage{amssymb,amsfonts,amsmath}
\usepackage{adjustbox}
\usepackage{hyperref}
\usepackage{braket}
\usepackage{color}
\usepackage{comment}

\begin{document}

\title{Two-atom-thin topological crystalline insulators\\lacking out of plane inversion symmetry}

\author{Salvador Barraza-Lopez}
\email{sbarraza@uark.edu}
\affiliation{Department of Physics, University of Arkansas, Fayetteville, Arkansas 72701, United States}
\affiliation{MonArk NSF Quantum Foundry, University of Arkansas, Fayetteville, Arkansas 72701, United States}
\author{Gerardo G. Naumis}
\email{naumis@fisica.unam.mx}
\affiliation{Instituto de F\'isica, Universidad Nacional Aut\'onoma de M\'exico, Mexico City, Mexico}

\begin{abstract}
A two-dimensional topological crystalline insulator (TCI) with a single unit cell (u.c.) thickness is demonstrated here. To that end, one first shows that tetragonal ($C_4$ in-plane) symmetry is not a necessary condition for the creation of zero-energy metallic surface states on TCI slabs of finite-thicknesses, because zero-energy states persist even as all the in-plane rotational symmetries--furnishing topological protection--are completely removed. {In other words, zero-energy levels on the model are not due to (nor are they protected by) topology.} Furthermore, effective twofold energy degeneracies taking place at {few discrete $k-$points away from zero energy} in the bulk Hamiltonian{--that are topologically protected--}persist at the u.c.~thickness limit. The chiral nature of the bulk TCI Hamiltonian permits creating a $2\times 2$ {\em square} Hamiltonian, whose topological properties {\em remarkably hold invariant at both the bulk and at the single u.c.~thickness limits}. The identical topological characterization for bulk and u.c.-thick phases is further guaranteed by a calculation involving Pfaffians. This way, a two-atom-thick TCI is deployed hereby, in a demonstration of a topological phase that holds both in the bulk, and in two dimensions.
\end{abstract}

\maketitle


\section{Introduction}

Studies of the topology of the electronic band structure started by describing the quantum Hall effect according to the Chern number \cite{TKNN}, and continued with the prediction of a quantum spin Hall (QSH) effect \cite{KANE1,KANE2}. { Several} QSH insulators were experimentally observed in mercury telluride quantum wells \cite{BHZ,KONIG07}, Bi$_{1-x}$Sb$_x$ alloys \cite{HSIEH08}, and bismuth chalcogenides \cite{ZHANG09}, and these materials are now commonly known as {\em strong topological insulators} (TIs) \cite{REV1,REV2,REV3}. The topology of the electronic band structure of {\em strong} TIs guarantees the existence of {robust surface conducting states} in what is known as the {\em bulk-boundary correspondence} \cite{REV1,REV2,REV3,Bernevig2013,Asboth_2016}, and time-reversal symmetry (TRS) produces a Kramers degeneracy on strong TIs. TRS is antiunitary, and customarily written as the product of a unitary operator and the complex conjugation operator \cite{Fruchart_2013,Lau2018}.  While 2D materials such as stanene are predicted to be 2D TIs \cite{PhysRevLett.111.136804} and Bi$_2$Se$_3$ \cite{REV1} is a 3D TI, it is rather unusual to find identical topological phases {\em on the same model or material platform } at both the 3D and 2D limits.

A TCI is a different kind of {{\em weak}} topological insulating phase that is protected by crystalline symmetries \cite{Fu}; a spinless electronic system in which the product of a crystal symmetry and time reversal symmetry produces an effective Kramers degeneracy, and whose characterization \cite{Fu} relies on integrals carried out over {\em open paths} in momentum space. The TCI phase has been observed on the {(high-temperature) cubic phase of SnTe \cite{HSIEH12}, which becomes a ferroelectric (not-TCI) phase at the unit cell thickness limit \cite{kai2016,RevModPhys.93.011001}}. {Not relying on spin but rather on crystalline symmetries, TCIs present robust edge states against magnetic impurities  \cite{Liu_2020}.} A TCI phase has been predicted on {\em inversion symmetric} SnTe films \cite{Fu1,Fu2,Fu3} with a thickness larger than a single unit cell. {The mirror reflection symmetry of Pb$_{1-x}$Sn$_x$Se TCIs was broken by strain \cite{Okada_2013}, imparting mass to the otherwise massless Dirac fermions. In some TCIs, pressure induces a  closing and subsequent reopening of the band gap \cite{Rajaji_2022}. Recent first-principles calculations and symmetry analyses predicted that atomically-thin transition metal dichalcogenides with a large band gap are higher-order TCIs \cite{Qian_2022}. Magnonic \cite{Kondo_2021} and photonic analogs have been proposed to mimic the properties of atomic TCIs \cite{Liu2021,PhysRevB.106.035155}. Using a photonic analogue, fractional charge and  localized states were observed at disclinations in TCI phases \cite{Liu2021}, and correlated electronic states were studied on TCIs \cite{Manghi_2021} as well.}

{Here, we analyze the {\em original} TCI model, for which no dedicated studies at the unit cell thickness limit have been carried out yet. Indeed, Fu’s original publication introducing TCIs deals with a {\em semi-infinite slab} as the band structure of one of the two surfaces is not visible on the slab electronic dispersion [compare Fig.~2(b) on Ref.~\cite{Fu} to Fig.~3(b) on Ref.~\cite{Manghi_2021}--which only published last year, and which does contain the dispersion of a {\em finite-size slab}--to verify our assertion]. This realization leaves a door for the study of topological properties of two-dimensional analogs of Ref.~\cite{Fu} wide-open; something that has not been done yet despite of that original's work longevity.} This opportunity came to our attention while studying the properties of {1-unit-cell-thick, {\em ferroelectric}} SnTe \cite{kai2016,KaiAdvMat,PRBKaloni,PhysRevLett.122.206402,RevModPhys.93.011001}.

We determine that the topological invariant characterizing the square of the {original} TCI Hamiltonian in Ref.~\cite{Fu}--one {\em lacking inversion symmetry with respect to the $x-y$ plane}--does not change its form in going from the bulk (3D) onto a u.c.-thick slab that still maintains the crucial $C_4$ in-plane rotational symmetry. Furthermore, if one accepts a definition of a topological invariant as $(-1)^{\nu_{\bar{\Gamma}\bar{M}}}$ at the 2D limit (because the $Z$ and $A$ points in momentum space lack meaning in 2D), such invariant remains negative and hence the TCI retains its topological properties when it is one u.c.~thick. These {unexpected} results establish that the weak topology of the original TCI phase stands unaffected ({\em i.e.}, that it persists) across spatial dimensions.

The following program is developed to that end:
The {\em original} TCI Hamiltonian, and an analysis of hopping in terms of Slater-Koster integrals \cite{Slater} are provided in Sec.~\ref{sec:1a}. The effective Kramers degeneracy, and the chiral nature of the Hamiltonian [which gives rise to a block diagonal square Hamiltonian], are discussed in Sec.~\ref{sec:1b}.
 An analysis of the electronic band structure of finite-size slabs is presented in Sec.~\ref{sec:1d}. The twofold energy degeneracy of pairs of bands at the $\bar{\Gamma}$ and $\bar{M}$ points is emphasized, as well as the nodal nature of zero-energy surface states. Analytical low-energy dispersions about the $\bar{M}$ point are contributed there as well.
 Then, a Hamiltonian without in-plane symmetry is developed in Sec.~\ref{sec:1e} to show how effective energy degeneracies underpinning the TCI are lifted by the reduction of symmetry. Twofold degenerate states at the $\bar{\Gamma}$ and $\bar{M}$ points persist down to the u.c.~thickness limit due to the block-diagonal nature of the Hamiltonian at these high-symmetry points, in which the $p_x$ and $p_y$ states become decoupled. It is {\em explicitly} shown here that the bulk-boundary correspondence does not hold for the TCI: zero-energy surface states on exposed surfaces of sufficiently thick slabs persist even as the crystal symmetry is reduced and the effective degeneracy underpinning the TCI phase does not exist any longer.
 From this point on, the focus returns to the $C_4$ in-plane symmetric TCI phase. We determine in Sec.~\ref{sec:I} that the TCI Hamiltonian \cite{Fu} is a chiral square-root representation of another $2\times 2$ Hamiltonian \cite{PhysRevB.95.165109}, whose vector field (an indicator of topology \cite{Asboth_2016}) does not change its form in the bulk and the single u.c.~thickness limits.
 The work ends with an explicit calculation of Pfaffians at the bulk and 2D limits in Sec.~\ref{sec:III}, where an identical topological characterization--as already obtained from a squared Hamiltonian--holds in 3D and 2D, for a {\em double verification} of the main claim made on this manuscript.
Conclusions are provided in Sec.~\ref{sec:IV}.


\section{Original TCI Hamiltonian and Slater-Koster analysis of its hopping terms}\label{sec:1a}
\begin{figure}[tb]
\begin{center}
\includegraphics[width=0.48\textwidth]{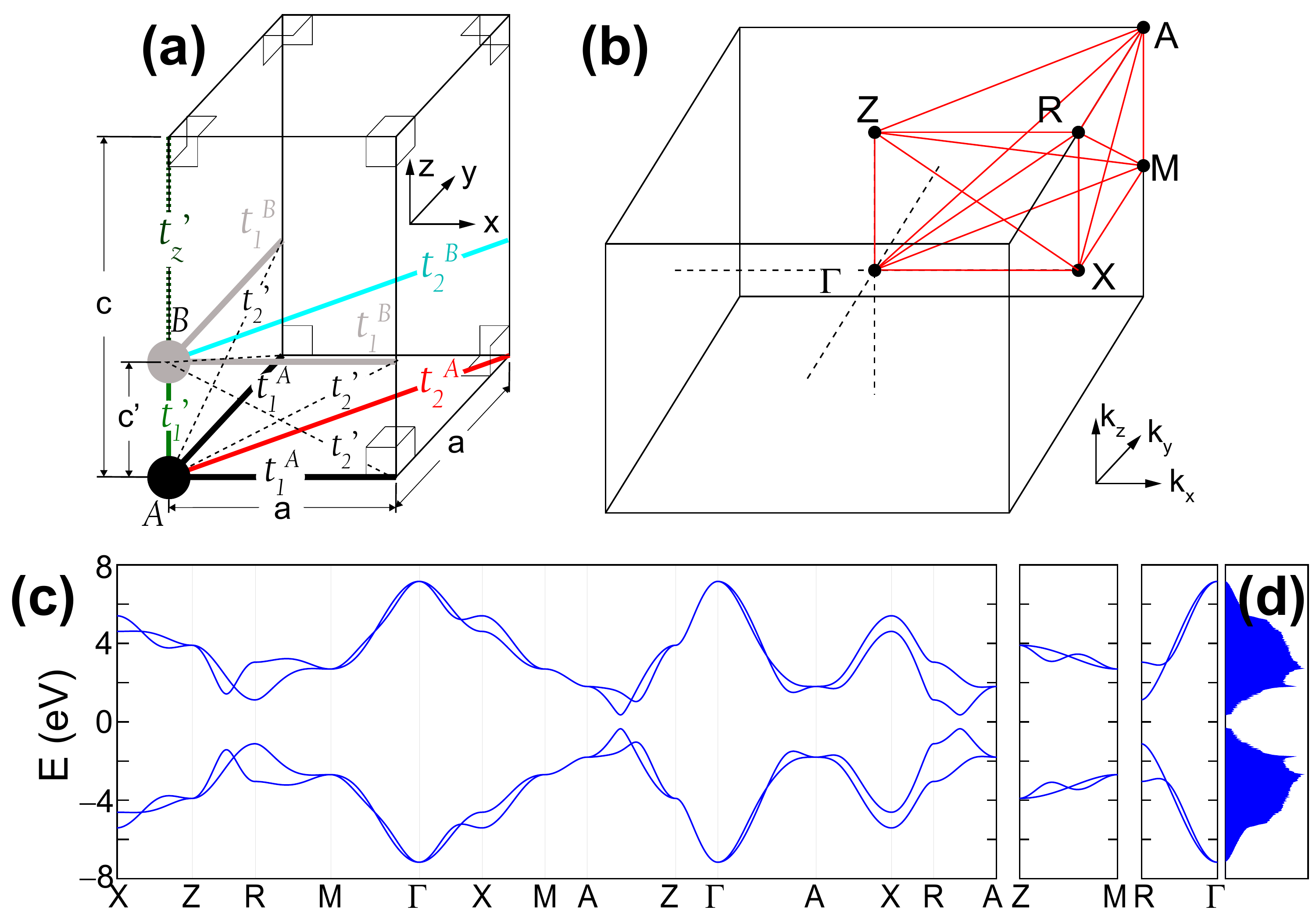}
\end{center}
\caption{(a) Tetragonal lattice (having a $C_4$ in-plane symmetry) with hopping terms indicated. (b) High-symmetry points and path in reciprocal space. (c) Electron-hole symmetric band structure of the bulk Hamiltonian; {  note the double degeneracy of states at $\Gamma$, $M$, $Z$, and $A$ high-symmetry points. (d) Density of states showcasing an $E_g=0.64$ eV bandgap.}\label{fig:F1}}
\end{figure}

Consider the four-by-four two-atomic-site bulk Hamiltonian containing one $p_x$ and one $p_y$ orbital per atomic site on the tetragonal lattice defined in Ref.~\cite{Fu}, on a u.c.~whose volume $a^2c$ is depicted on Fig.~\ref{fig:F1}(a), and in which hopping terms were indicated, too:
\begin{equation}\label{eq:Hbulk}
\begin{aligned}
H(\mathbf{k})=&
\left(
\begin{matrix}
H^A(k_x,k_y) & H^{AB}(\mathbf{k})\\
H^{AB\dagger}(\mathbf{k}) & H^B(k_x,k_y)\\
\end{matrix}
\right),
\end{aligned}
\end{equation}
with $\mathbf{k}=(k_x,k_y,k_z)$,
\begin{eqnarray}\label{eq:Halpha}
&H^\alpha(k_x,k_y)= 2t_1^\alpha
\left(
\begin{matrix}
\cos (k_xa) & 0\\
0 & \cos (k_ya)
\end{matrix}
\right)+\\
&2t_2^\alpha
\left(
\begin{matrix}
\cos (k_xa)\cos (k_ya) & \sin (k_xa)\sin (k_ya)\\
\sin (k_xa)\sin (k_ya) & \cos (k_xa)\cos (k_ya)
\end{matrix}
\right),\nonumber
\end{eqnarray}
($\alpha=A$, or $B$) and:
\begin{eqnarray}\label{eq:HAB}
H^{AB}(\mathbf{k})&=\left[t_1'+2t_2'(\cos (k_xa) +\cos (k_ya))+t_z'e^{ik_zc}\right]\sigma_0\nonumber\\
&=\left[f_s(k_x,k_y)+t_z'\text{e}^{ik_zc}\right]\sigma_0=f_b(\mathbf{k})\sigma_0,
\end{eqnarray}
where $b$ stands for {\em bulk}. ($H^{\alpha}(\mathbf{k})$ {is a real matrix which} only depends on $k_x$ and $k_y$ in Ref.~\cite{Fu}, and such dependency was made explicit in Eqn.~\eqref{eq:Halpha}.)

In Eqns.~\eqref{eq:Hbulk}, \eqref{eq:Halpha}, and \eqref{eq:HAB}, $\sigma_0$ is the 2$\times$2 identity matrix, $\dagger$ indicates the Hermitian adjoint operator, $t_1^A=-t_1^B=1$, $t_2^A=-t_2^B=0.5$, $t_1'=2.5$, $t_2'=0.5$, and $t_z'=2$ (eV units are adopted for the hopping terms for definiteness here). Figure \ref{fig:F1}(b) shows the $k-$point path used to plot the bulk band structure on Fig.~\ref{fig:F1}(c), and Fig.~\ref{fig:F1}(d) shows the density of states (DOS) sampled over the tetragonal first Brillouin zone.

We link the tight-binding Hamiltonian in Eqn.~\eqref{eq:Hbulk} to expressions provided by Slater and Koster involving $p_x$ and $p_y$ orbitals \cite{Slater}, to guide the design of low-symmetry Hamiltonians in Sec.~\ref{sec:1e}. {Leaving the details to the Supplementary Material,} $H^{(1)}_{pp\sigma}=t_1^{\alpha}$, $H^{(1)}_{pp\pi}=0$,  $H^{(2)}_{pp\sigma}=t_2^{\alpha}$, and $H^{(2)}_{pp\pi}=0$, where the upper index $^{(j)}$ stands for first ($^{(1)}$) or second ($^{(2)}$) nearest neighbor.

As for out-of-plane second nearest neighbors, the $p_x^Ap_y^B$ entry is equal to zero, while the $p_x^Ap_x^B$ and $p_y^Ap_y^B$ entries become:
\begin{eqnarray}\label{eq:HppsHppp}
& 2\left[ \frac{a^2}{a^2+c'^2}H^{'(2)}_{pp\sigma} +\frac{c'^2}{a^2+c'^2}H^{'(2)}_{pp\pi}\right]\cos(k_xa)+\nonumber\\
& 2H^{'(2)}_{pp\pi}\cos(k_ya),\text{ and}\\
& 2H^{'(2)}_{pp\pi}\cos(k_xa)+ \nonumber\\
& 2\left[ \frac{a^2}{a^2+c'^2}H^{'(2)}_{pp\sigma} +\frac{c'^2}{a^2+c'^2}H^{'(2)}_{pp\pi}\right]\cos(k_ya),\nonumber
\end{eqnarray}
respectively ($a$ and $c'$ were defined on Fig.~\ref{fig:F1}(a)). According to Eqn.~\eqref{eq:HAB}, the term $t_2'$ multiplying $\cos(k_xa)+\cos(k_ya)$ is identical for $p_x^Ap_x^B$ and for $p_y^Ap_y^B$, which sets a constraint $H^{'(2)}_{pp\sigma}=H^{'(2)}_{pp\pi}=t_z'$ on Eqn.~\eqref{eq:HppsHppp}. The present analysis will be employed to design a Hamiltonian on a monoclinic unit cell later on.

\section{Effective Kramers degeneracy and chiral symmetry of the original TCI Hamiltonian}\label{sec:1b}

The $C_4$ symmetry within the $xy-$plane is represented by the $e^{i\sigma_y\pi/2}$ operator, and the rotation matrix for the bulk Hamiltonian is built out of two copies of it, {one per atom}:
\begin{equation}\label{eq:U}
\mathbb{U}=\begin{pmatrix}
 0 & 1 & 0 & 0\\
-1 & 0 & 0 & 0\\
 0 & 0 & 0 & 1\\
 0 & 0 &-1 & 0
\end{pmatrix},
\end{equation}
which has double degenerate $\pm i$ eigenvalues.

Imaginary contributions enter $H(\mathbf{k})$ through $t'_ze^{ik_z}\sigma_0$ in Eqn.~\eqref{eq:HAB}. Nevertheless, this phase is real in the $k_z=0$ and $k_z=\pi$ planes {(where Pfaffians will be computed \cite{Fu})}, making the Hamiltonian real there as well. $H(k_x,k_y,0)$ and $H(k_x,k_y,\pi/c)$ thus satisfy an effective Kramers degeneracy:
\begin{equation}\label{eq:Kramers}
H(\mathbf{k})=\Xi H(\mathbf{k})\Xi,\text{ and }\Xi|u_m(\mathbf{k})\rangle=-|u_m(\mathbf{k})\rangle
\end{equation}
at these two planes, with $\Xi=\mathbb{U}^2T$, $m=1,2$, and $T$ being the antiunitary time reversal operator (a mere complex conjugation for the spinless fermions being considered here).

\begin{figure}[tb]
\begin{center}
\includegraphics[width=0.42\textwidth]{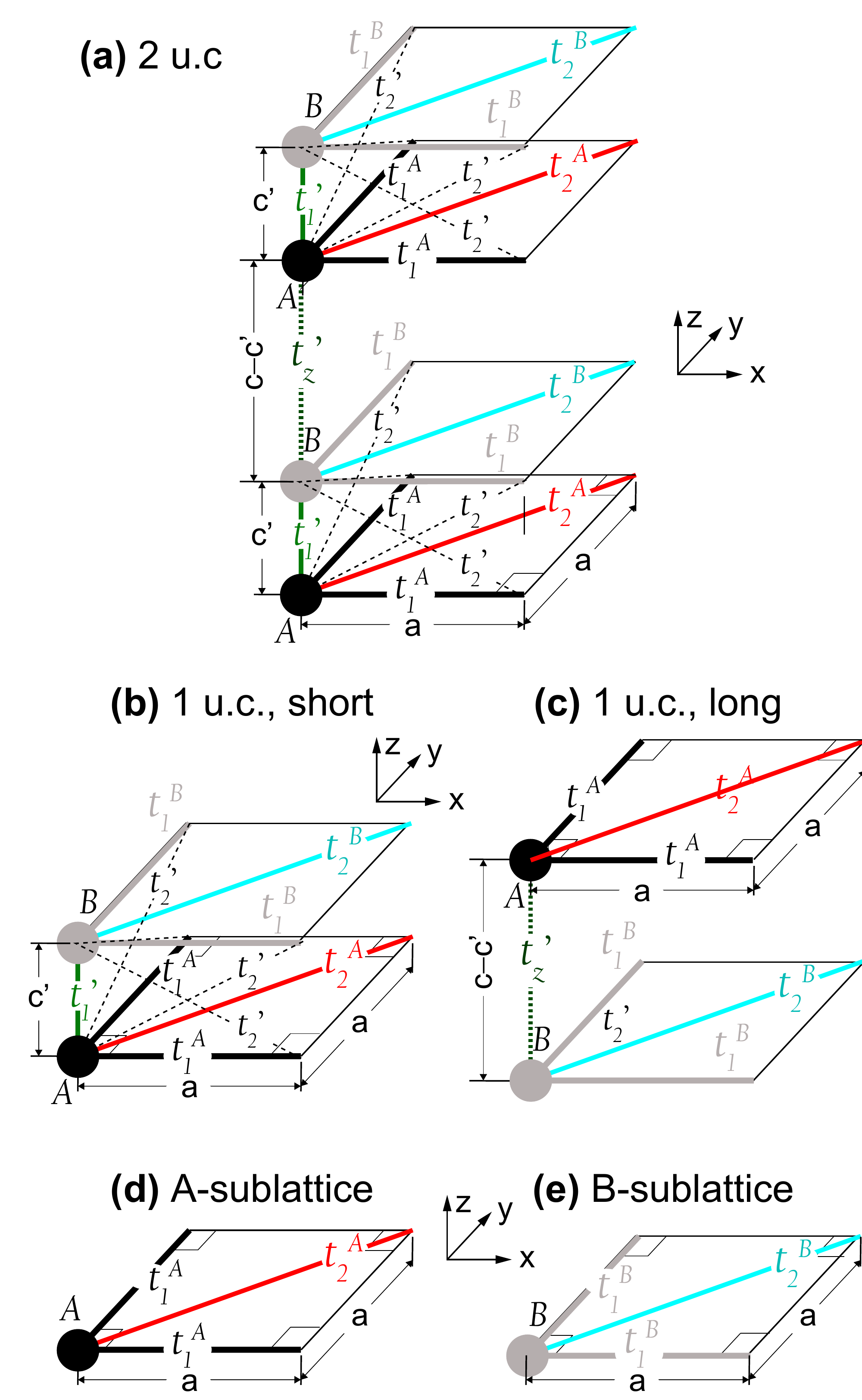}
\end{center}
\caption{Atomistic structure and schematics of hopping terms for selected slabs: (a) Two u.c.s. (b) ``Short'' u.c. (c) ``Long'' u.c. (d) A-sublattice and (e) B-sublattice slabs. Slabs are periodic on the $x-$ and $y-$directions.\label{fig:newF3}}
\end{figure}

\begin{figure*}
\begin{center}
\includegraphics[width=0.96\textwidth]{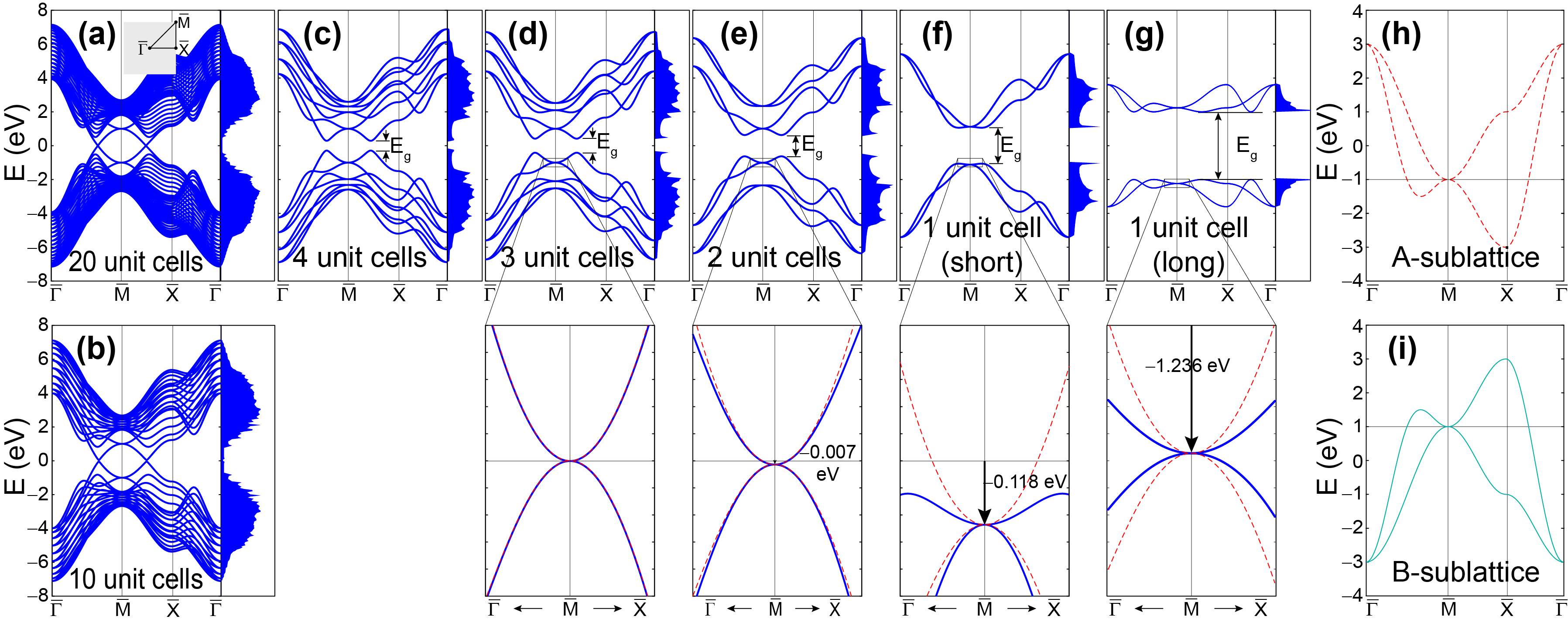}
\end{center}
\caption{Electronic dispersion of slabs containing $n$ u.c.s: (a) $n=20$, (b) $n=10$, (c) $n=4$, (d) $n=3$, (e) $n=2$, (f) $n=1$, short u.c., and (g) $n=1$, long u.c. {As illustrated on Fig.~\ref{fig:newF3}, these slabs have atoms belonging to opposite sublattices facing vacuum.} (h) Electronic dispersion for a slab with an $A$ sublattice thickness, { and (i) with an $B$ sublattice thickness only}. Orthogonality of states cannot be guaranteed away from the $\bar{\Gamma}$ and $\bar{M}$ points, leading to energy gaps $E_g$ at zero energy that are confirmed by DOS calculations. Lower subplots are zoom-ins showing overlap with the dispersion of the $A-$sublattice (similar behavior exists for the dispersion around $\bar{M}$ at 1 eV). Two-fold degenerate energy crossings never split at the $\bar{\Gamma}$ and $\bar{M}$ points, regardless of slab thickness.\label{fig:Fi3}}
\end{figure*}

The choice  $t_1^A=-t_1^B$ and $t_2^A=-t_2^B$ ({{\em i.e.}, $H^B(k_x,k_y)=-H^A(k_x,k_y)$}) creates a {\em chiral symmetry} that makes the dispersion {on Eqn.~\eqref{eq:Hbulk}} electron-hole symmetric:
\begin{equation}\label{eq:H2reduced}
H(\mathbf{k})=\left( \begin{array}{cc}
H^{A}(k_x,k_y) & H^{AB}(\mathbf{k}) \\
H^{AB\dagger}(\mathbf{k})  & -H^{A}(k_x,k_y)  \\
\end{array} \right).
\end{equation}

The chirality of the bulk Hamiltonian, Eqn.~\eqref{eq:H2reduced}, implies that for any eigenvector with energy $E$: $|\boldsymbol{\Psi}_E((\mathbf{k}))\rangle=(\psi_{p_x}^A,\psi_{p_y}^A,\psi_{p_x}^B,\psi_{p_y}^B)^T$, there exists an eigenvector with energy $-E$: $|\boldsymbol{\Psi}_{-E}((\mathbf{k}))\rangle=(\psi_{p_x}^A,\psi_{p_y}^A,-\psi_{p_x}^B,-\psi_{p_y}^B)^T$ \cite{Naumis_Binary}. The electron-hole symmetry of $H(\mathbf{k})$ is evident on Figs.~\ref{fig:F1}(c) and \ref{fig:F1}(d), and it will lead to a block-diagonal square Hamiltonian in Sec.~\ref{sec:I}.

When applied to the bulk Hamiltonian [Eqn.~\eqref{eq:H2reduced}], Eqn.~\eqref{eq:U} satisfies the symmetry requirements stated as Eqn.~(4) in Ref.~\cite{Fu}. In addition,
\begin{equation}\label{eq:UH}
[\mathbb{U},H(\mathbf{k})]=0,
\end{equation}
at $\Gamma$, $M$, $Z$, and $A$. { Eqn.~\eqref{eq:UH} is crucial to ensure two-fold degenerate eigenvalues at these high-symmetry points. The energy degeneracies at these high-symmetry points are the crucial ingredient for the non-trivial topology of the original TCI model.}


\section{Electronic structure of slab Hamiltonians from the original TCI model}\label{sec:1d}
{As indicated at the introduction,} Ref.~\cite{Fu} emphasizes surface states on a {\em semi-infinite} slab configuration. {Published only last year,} Ref.~\cite{Manghi_2021} shows the electronic structure of a {\em finite} slab with surface terminations on opposite sublattices {for the first time}. Here, {\em we aim to explore the topology {of the original TCI model} right at the 2D limit}, and study slabs with varying thicknesses next with that goal in mind. To that end, {Fig.~\ref{fig:newF3}(a) exemplifies a slab with two u.c. thickness, while Figs.~\ref{fig:newF3}(b-c) show slabs with one u.c.~thickness, and Figs.~\ref{fig:newF3}(d-e) represent slabs with sublattice thickness.}  Fig.~\ref{fig:Fi3} shows the electronic dispersion for TCI slabs with (a) twenty, (b) ten, (c) four, (d) three, (e) two, and (f-g) one u.c.s.

{To be explicit, all Hamiltonians arising from the structures shown on Fig.~\ref{fig:newF3} are:
\begin{eqnarray}\label{eq:H2uc}
H_{2uc}(k_x,k_y)=\\
\left(\begin{smallmatrix}
H^A(k_x,k_y)         & f_s(k_x,k_y)\sigma_0 &   \mathbb{O}         & \mathbb{O}\\
f_s(k_x,k_y)\sigma_0 &-H^A(k_x,k_y)         & t_z'\sigma_0         & \mathbb{O}\\
\mathbb{O}           & t_z'\sigma_0         & H^A(k_x,k_y)         & f_s(k_x,k_y)\sigma_0\\
\mathbb{O}           & \mathbb{O}           & f_s(k_x,k_y)\sigma_0 &-H^A(k_x,k_y)
\end{smallmatrix}\right)\nonumber
\end{eqnarray}
for the two-u.c.~structure shown as Fig.~\ref{fig:newF3}(a), where no complex conjugation was used--as all matrices involved are real--and $\mathbb{O}$ is a $2\times 2$ matrix full of zeroes. The eigenvalues of $H_{2uc}(k_x,k_y)$ are displayed on Fig.~\ref{fig:Fi3}(e).}

{As illustrated on Fig.~\ref{fig:newF3}(b-c),} there are two different ways to build a slab with a single u.c.~thickness. On the first choice (labeled {\em short}), the $A$ and $B$ atoms are separated by a distance $c'$ [see Figs.~\ref{fig:F1}(a) {and \ref{fig:newF3}(a)}]. In the second choice (dubbed {\em long}) the $A$ atom is placed above the $B$ atom at a height $c-c'$ {[Fig.~\ref{fig:newF3}(b)]}. The two Hamiltonians for these slabs are:
\begin{equation}\label{eq:2DC4s}
H_s(k_x,k_y)=
\begin{pmatrix}
H^A(k_x,k_y) & f_s(k_x,k_y)\sigma_0\\
f_s(k_x,k_y)\sigma_0 &-H^A(k_x,k_y)
\end{pmatrix}
\end{equation}
for the short u.c.~slab [Fig.~\ref{fig:newF3}(b)], and:
\begin{equation}\label{eq:2DC4l}
H_l(k_x,k_y)=
\begin{pmatrix}
H^A(k_x,k_y) & t_z'\sigma_0\\
t_z'\sigma_0 &-H^A(k_x,k_y)
\end{pmatrix}
\end{equation}
for the long u.c.~slab [Fig.~\ref{fig:newF3}(b)]. These slabs with one u.c.~thickness [Fig.~\ref{fig:Fi3}(f) {\em or} ~\ref{fig:Fi3}(g)] are semiconductors with $E_g=2.122$ eV or $E_g=3.956$ eV energy band gaps, respectively. {No investigations of topological properties of those u.c.-thick slabs have ever been reported in the literature.}

Figures \ref{fig:Fi3}(h) and \ref{fig:Fi3}(i) depict the electronic dispersions of slabs having a sublattice thickness, {whose Hamiltonians are either $H^{A}(k_x,k_y)$, or $-H^{A}(k_x,k_y)$}, respectively {[see Figs.~\ref{fig:newF3}(d-e) for schematics], and which will be useful to fit electronic dispersions around the $\bar{M}$ point.
Building slabs with increased thickness is relatively straightforward. Defining
\begin{equation}\label{eq:tau}
\tau=
\left(
\begin{matrix}
\mathbb{O}   & \mathbb{O}\\
t_z'\sigma_0 & \mathbb{O}
\end{matrix}
\right),
\end{equation}
(which remains a real matrix),  Eqn.~\eqref{eq:H2uc} can be rewritten as:
\begin{eqnarray}\label{eq:H2ucv2}
H_{2uc}(k_x,k_y)=
\begin{pmatrix}
H_s(k_x,k_y) &\tau \\
\tau^T & H_s(k_x,k_y)
\end{pmatrix}
\end{eqnarray}
where $\tau^T$ is the transpose of $\tau$. The plots shown on Figs.~\ref{fig:Fi3}(a) through \ref{fig:Fi3}(d) are the eigenvalues of Hamiltonians built in a way similar to Eqn.~\eqref{eq:H2ucv2}. For instance, \ref{fig:Fi3}(c) shows the eigenvalues of
\begin{eqnarray}\label{eq:4ucs}
H_{4uc}(k_x,k_y)=\\
\left(
\begin{smallmatrix}
  H_s(k_x,k_y) & \tau         & \mathcal{O}  & \mathcal{O} \\
  \tau^T       & H_s(k_x,k_y) & \tau         & \mathcal{O} \\
  \mathcal{O}  & \tau^T       & H_s(k_x,k_y) & \tau        \\
  \mathcal{O}  & \mathcal{O}  & \tau^T       & H_s(k_x,k_y)
\end{smallmatrix}
\right),\nonumber
\end{eqnarray}
where $\mathcal{O}$ is a $4\times 4$ matrix full of zeroes.}

Horizontally scaled DOS plots confirm the existence of the electronic bandgaps $E_g$ as observed in the band structure plots. {[Explicit code for calculating band structures and DOS for slabs containing ten and two unit cells is facilitated as Supporting Material. We also provide a study of the difference among hybridized states near zero energy, and the block-diagonal nature of slab Hamiltonians at high-symmetry points there.]} With the exception of Figs.~\ref{fig:Fi3}(h) and \ref{fig:Fi3}(i) (which are slabs made out of only one sublattice, and whose electronic occupation differs at the $\bar{\Gamma}$ and $\bar{M}$ points), all slabs {leading to the band structures on} Fig.~\ref{fig:Fi3} have opposite sublattice atoms at each of their two exposed surfaces {\em and hence lack inversion symmetry with respect to the plane defining those exposed surfaces}. {[The electronic dispersion for slabs having atoms of the same sublattice facing vacuum can also be found as Supporting Material.]}

There is always a region in $k-$space centered about $\bar{M}$ in which the surface dispersion is {\em  given analytically} by either $H^A(k_x,k_y)$ (lower exposed, $A-$sublattice surface), or by $-H^A(k_x,k_y)$ (upper exposed, $B-$sublattice surface), so that the dispersion of the surface bands can be  known exactly; {such discussion can be found as Supporting Material.}

\begin{figure}[tb]
\begin{center}
\includegraphics[width=0.48\textwidth]{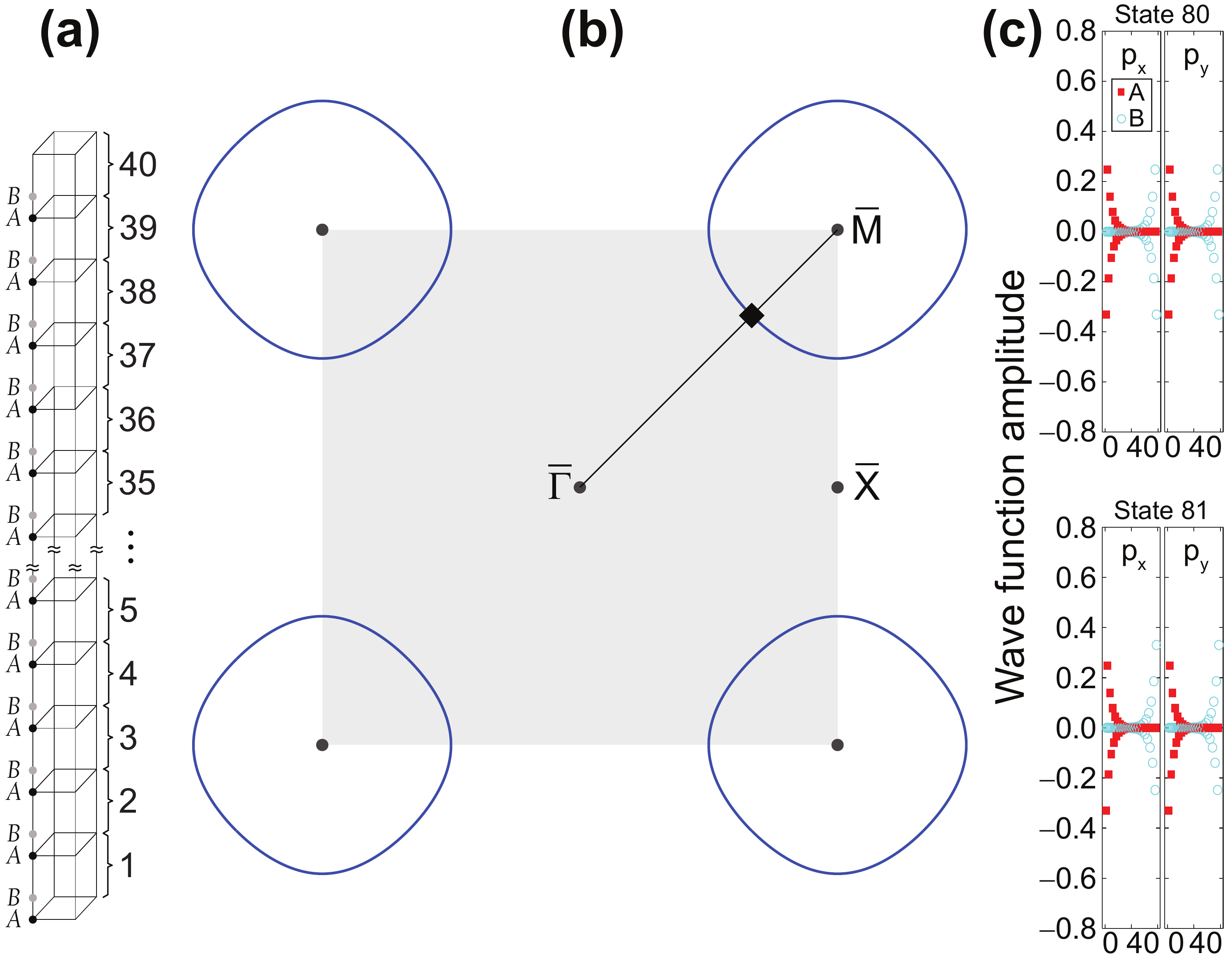}
\end{center}
\caption{{(a) Schematic depiction of a 40 u.c.-thick slab. (b)} Nodal structure (blue curves) of the Fermi energy for this slab. (c) Zero energy states at the crossing marked by a rotated square in panel (b), located at $0.673\bar{M}$.\label{fig:FS}}
\end{figure}

The electronic dispersion at zero energy will be discussed in Sec.~\ref{sec:1e}, making it necessary to identify the points in reciprocal space leading to zero energy states on sufficiently thick slabs. Figure \ref{fig:FS}(a) contains a depiction of zero-energy electronic dispersion contours for a slab with a thickness of 40 u.c.s. The electronic bandgaps to be shown on Fig.~\ref{fig:newMonoclinic} were obtained along the $\bar{\Gamma}-\bar{M}$ line [or the $\bar{\Gamma}-\bar{C}$ line for an oblique lattice]. Figure \ref{fig:FS}(b) verifies that states at zero energy are surface states.


\section{3D and 2D Hamiltonians arising from the original TCI model but with reduced symmetries}\label{sec:1e}

As indicated in Ref.~\cite{RevModPhys.88.035005}, ``the absence of gapless modes at boundaries that break the spatial symmetries does not indicate trivial bulk topology and therefore cannot be used to infer the topology of TCIs.'' In other words, the bulk-boundary correspondence (which applies to strong TIs \cite{REV1,REV2,REV3}) does not hold for (weak) TCIs, and {\em an explicit proof}--which we have not found in the literature--is provided in what follows.

The proof requires showing that {\em zero-energy states}--that develop on sufficiently thick slabs--persist even when $C_4$ symmetry is removed, so that topological protection does not exist anymore. In pursuing the proof, one also sees the effective degeneracy being lifted at four high-symmetry points on the bulk first Brillouin zone (and, accordingly, on two $k-$points on the 2D Brillouin zone) when $C_4$ symmetry no longer holds. The link among $C_4$ symmetry and the effective degeneracy seen at the $\Gamma$ and $M$ points in the bulk persists at the $\bar{\Gamma}$ and $\bar{M}$ points in 2D.

\begin{figure}[tb]
\begin{center}
\includegraphics[width=0.48\textwidth]{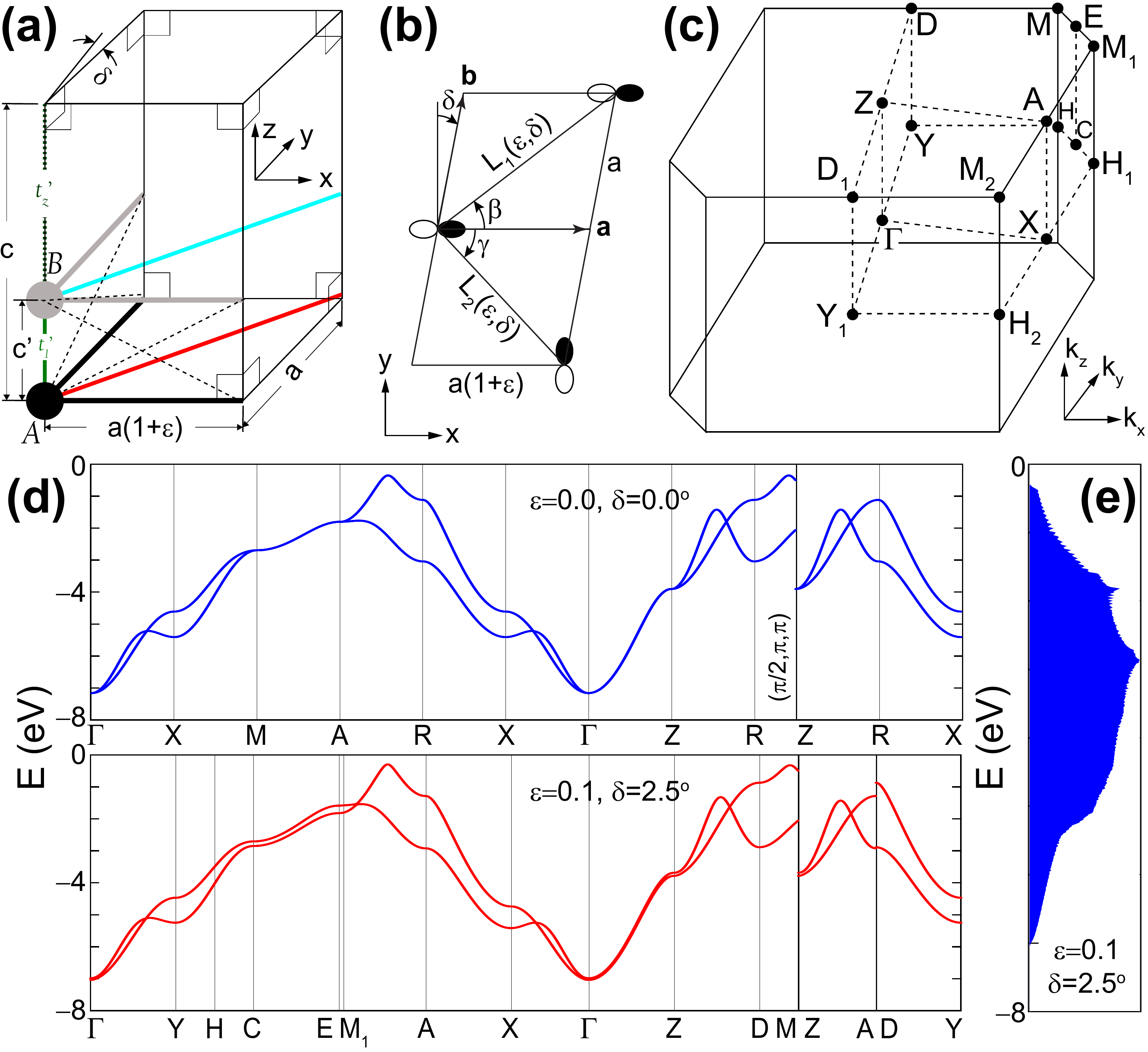}
\end{center}
\caption{(a) Monoclinic lattice (having no in-plane rotational symmetries). (b) The angle for second-nearest-neighbor hopping terms changes from $45^{\circ}$ to either $\beta$ or $\gamma$; one diagonal becomes larger ($L_1$), while another ($L_2$) shortens. (c) High-symmetry points and path in reciprocal space. (d) Band structure of the bulk Hamiltonian for occupied states (the band structure remains electron-hole symmetric).  Twofold energy degeneracies at $\Gamma$, $S$, $Z$, and $R$ high-symmetry points become lifted as the $C_4$ in-plane symmetry ($\epsilon=0.0$ and $\delta=0.0^{\circ}$; upper subplot) is removed altogether ($\epsilon=0.1$ and $\delta=2.5^{\circ}$; lower subplot). (e) DOS showcasing an $E_g=0.56$ eV bandgap for the monoclinic structure with $\epsilon=0.1$ and $\delta=2.5^{\circ}$.\label{fig:monoclinic}}
\end{figure}

For this purpose, Fig.~\ref{fig:monoclinic}(a) depicts a monoclinic unit cell created by lattice vectors $\mathbf{a}=a(1+\epsilon,0,0)$, $\mathbf{b}=a(\sin\delta,\cos\delta,0)$, and $\mathbf{c}=c(0,0,1)${, where the two parameters $\epsilon$ and $\delta$ bring the unit cell away from a tetragonal structure onto the monoclinic one. As seen on Fig.~\ref{fig:monoclinic}(b), $\epsilon>0$ represents an elongation of one in-plane lattice vector ($\mathbf{a}$), while $\delta>0$ indicates a tilt of the second in-plane lattice vector $\mathbf{b}$ onto $\mathbf{a}$.}

{Non-zero values for $\epsilon$ and $\delta$ alter Slater-Koster integrals involved in hopping. For small values of $\epsilon$, we introduce a reduction of $t_1^{\alpha}$ and of $t_2'$ along the $x-$direction by $1-\epsilon$.} In-plane second nearest neighbor distances {(whose original magnitude is $\sqrt{2}a$)} get modified under the monoclinic symmetry as well [see Fig.~\ref{fig:monoclinic}(b)]:
\begin{eqnarray}
L_1(\epsilon,\delta)&=\sqrt{2}a\sqrt{(1+\epsilon)(1+\sin\delta)+\epsilon^2/2},\text{ and}\nonumber\\
L_2(\epsilon,\delta)&=\sqrt{2}a\sqrt{(1+\epsilon)(1-\sin\delta)+\epsilon^2/2},
\end{eqnarray}
{leading to additional renormalization of hopping integrals. Angles among in-plane second nearest neighbors change from $45^{\circ}$ on the tetragonal lattice onto}:
\begin{eqnarray}
\cos\beta (\epsilon,\delta)&=\sqrt{1-\frac{a^2\cos^2\delta}{(L_1(\epsilon,\delta))^2}},\text{ and}\nonumber\\
\cos\gamma(\epsilon,\delta)&=\sqrt{1-\frac{a^2\cos^2\delta}{(L_2(\epsilon,\delta))^2}},
\end{eqnarray}
and induce additional renormalization of hopping integrals.

Figure \ref{fig:monoclinic}(c) depicts the first Brillouin zone for the monoclinic lattice whose high symmetry points were labeled following Ref.~\cite{Curtaolo}, and the monoclinic bulk Hamiltonian{
\begin{equation}\label{eq:Hmbulk}
H_m(\mathbf{k})=
\begin{pmatrix}
H_m^{A}(k_x,k_y)     & H_m^{AB}(\mathbf{k})\\
H_m^{AB\dagger}(\mathbf{k}) &-H_m^{A}(k_x,k_y)
\end{pmatrix}
\end{equation}
} {takes the following explicit form}:
\begin{widetext}
\begin{eqnarray}\label{eq:Bulkmonoclinic}
&H_m^{A}(k_x,k_y)=\\
&2t_1^{A}\left(
\begin{smallmatrix}
(1-\epsilon)\cos\left( k_xa(1+\epsilon) \right)+\sin^2\delta \cos\left( a(\sin\delta k_x+\cos \delta k_y) \right) & \frac{1}{2}\sin(2\delta)\cos\left( a(\sin\delta k_x+\cos \delta k_y) \right)\\
\frac{1}{2}\sin(2\delta)\cos\left( a(\sin\delta k_x+\cos \delta k_y) \right) & \cos^2\delta \cos\left( a(\sin\delta k_x+\cos \delta k_y) \right)
\end{smallmatrix}
\right)+
\left(
\begin{smallmatrix}
H^{A}_{m,p_x,p_x}             & H^{A}_{m,p_x,p_y}\\
H^{A}_{m,p_x,p_y}{}^{\dagger} & H^{A}_{m,p_y,p_y}
\end{smallmatrix}
\right)\nonumber
\end{eqnarray}
\end{widetext}
where
\begin{eqnarray*}
&H^{A}_{m,p_x,p_x}=\\
&\frac{2t_2^{A}\sqrt{2}a}{L_1(\epsilon,\delta)}\cos^2\beta \cos \left[ a(1+\epsilon+\sin\delta)k_x+a\cos\delta k_y \right]\\
&                      +\frac{2t_2^{A}\sqrt{2}a}{L_2(\epsilon,\delta)}\cos^2\gamma\cos \left[ a(1+\epsilon-\sin\delta)k_x-a\cos\delta k_y \right],
\end{eqnarray*}
\begin{eqnarray*}
&H^{A}_{m,p_x,p_y}=\\
&\frac{2t_2^{A}\sqrt{2}a}{L_1(\epsilon,\delta)}\cos\beta \sin\beta \cos \left[ a(1+\epsilon+\sin\delta)k_x+a\cos\delta k_y \right]\\
&                      -\frac{2t_2^{A}\sqrt{2}a}{L_2(\epsilon,\delta)}\cos\gamma\sin\gamma\cos \left[ a(1+\epsilon-\sin\delta)k_x-a\cos\delta k_y \right],
\end{eqnarray*}
and
\begin{eqnarray*}
&H^{A}_{m,p_y,p_y}=\\
&\frac{2t_2^{A}\sqrt{2}a}{L_1(\epsilon,\delta)}\sin^2\beta \cos \left[ a(1+\epsilon+\sin\delta)k_x+a\cos\delta k_y \right]\\
&                      +\frac{2t_2^{A}\sqrt{2}a}{L_2(\epsilon,\delta)}\sin^2\gamma\cos \left[ a(1+\epsilon-\sin\delta)k_x-a\cos\delta k_y \right].
\end{eqnarray*}

The $(1-\epsilon)$ prefactor seen on one term in Eqn.~\eqref{eq:Bulkmonoclinic} represents the decrease on the coupling strength {along the $x-$direction of the original hopping term $t_1^A$} due to the elongation of lattice vector $\mathbf{a}$. 
The modification of the hopping terms {$t_2^A$} on the second matrix on Eqn.~\eqref{eq:Bulkmonoclinic}--representing the interactions among in-plane second nearest neighbors--is captured by the ratio $\frac{\sqrt{2}a}{L_1(\epsilon,\delta)}$ or $\frac{\sqrt{2}a}{L_2(\epsilon,\delta)}$, which are slightly larger or smaller from unity, respectively. {A detailed derivation of the in-plane components of the monoclinic Hamiltonian is presented in the Supplementary Material.}

The matrix coupling electrons among the $A$ and $B$ sublattices in the monoclinic lattice looks as follows:
\begin{eqnarray}
H_m^{AB}(\mathbf{k})=
\left[t_1'+2t_2'(1-\epsilon)\cos\left( k_xa(1+\epsilon) \right)\right]\sigma_0\nonumber\\
+
\left[2t_2'\cos\left( k_xa\sin\delta +k_ya\cos\delta \right) + t_z'\text{e}^{ik_zc}
\right]\sigma_0,
\end{eqnarray}
as, once again, only the bond with a non-zero projection onto the $x-$axis became elongated {[leading to a single modified hopping $t_2'\to t_2'(1-\epsilon)$ along the $x-$direction]}. The band structure for the bulk is shown on Fig.~\ref{fig:monoclinic}(d), and the DOS  is depicted on Fig.~\ref{fig:monoclinic}(e) when $\epsilon=0.1$ and $\delta=2.5^{\circ}$. The breakdown of the energy degeneracy at high symmetry points is listed in Table \ref{ta:Tamonoclinic}.

{
Defining
\begin{eqnarray}
f_{s,m}(k_x,k_y)=t_1'+2t_2'(1-\epsilon)\cos(k_xa(1+\epsilon))\\
+2t_2'\cos(k_xa\sin \delta+k_ya\cos\delta)\nonumber
\end{eqnarray}
(which remains real),  and noticing that the vertical hopping $t_z'$ remain unaffected on the lowering of symmetry onto a monoclinic unit cell [such that $\tau$ in Eqn.~\ref{eq:tau} remains unchanged], we find the effect of the lowering of symmetry on slabs with 1 u.c.~thickness:
\begin{equation}\label{eq:smallm}
H_{s,m}(k_x,k_y)=
\begin{pmatrix}
  H_m^A(k_x,k_y) & f_{s,m}(k_x,k_y)\sigma_0 \\
  f_{s,m}(k_x,k_y)\sigma_0 & -H_m^A(k_x,k_y)
\end{pmatrix},
\end{equation}
for the small u.c., or
\begin{equation}\label{eq:longm}
H_{l,m}(k_x,k_y)=
\begin{pmatrix}
  H_m^A(k_x,k_y) & t_z'\sigma_0 \\
  t_z'\sigma_0 & -H_m^A(k_x,k_y)
\end{pmatrix},
\end{equation}
for the long u.c. A monoclinic 3D lattice becomes oblique in two-dimensions, and}
the breakdown of the effective degeneracy at the 2D limit for the oblique unit cell is demonstrated in Tables \ref{ta:Tamonoclinicshort} and \ref{ta:Tamonocliniclong} for both short and long u.c.s, respectively.

\begin{table}[tb]
\caption{Breakdown of the effective degeneracy of occupied states at high symmetry points on the (3D) monoclinic lattice, {Eqn.~\eqref{eq:Hmbulk}} (in eV). High symmetry points $H_2$ and $C$ turn into $M$, while k-points $M_2$ and $E$ turn into $A$ as the unit cell becomes tetragonal ($\epsilon=0.0$ and $\delta=0.0^{\circ}$).\label{ta:Tamonoclinic}}
\begin{tabular}{c|c|c|c}
\hline
\hline
         & $\epsilon=0.0$          & $\epsilon=0.1$          & $\epsilon=0.2$          \\
         & and $\delta=0.0^{\circ}$& and $\delta=2.5^{\circ}$& and $\delta=5.0^{\circ}$\\
\hline
$\Gamma$ & $-7.1589$, $-7.1589$    & $-7.0364$, $-6.9849$    & $-6.9209$, $-6.8221$\\
\hline
$H_2$    & $-2.6926$, $-2.6926$    & $-2.8618$, $-2.7235$    & $-3.0457$, $-2.8160$\\
\hline
$C$      & $-2.6926$, $-2.6926$    & $-2.8495$, $-2.7055$    & $-3.0014$, $-2.7535$\\
\hline
$Z$      & $-3.9051$, $-3.9051$    & $-3.7829$, $-3.6863$    & $-3.6740$, $-3.4843$\\
\hline
$M_2$    & $-1.8028$, $-1.8028$    & $-1.8050$, $-1.5765$    & $-1.7912$, $-1.3644$\\
\hline
$E$      & $-1.8028$, $-1.8028$    & $-1.8221$, $-1.5873$    & $-1.8461$, $-1.4077$\\
\hline
\hline
\end{tabular}\\
\end{table}

\begin{table}[tb]
\caption{Breakdown of the effective degeneracy of occupied states at high symmetry points on the oblique 2D lattice{--Eqn.~\eqref{eq:smallm}--}for the {\em short} u.c.~slab (in eV). High-symmetry $k-$points $\bar{H_2}$ and $\bar{C}$ turn into $\bar{M}$ as the unit cell becomes square ($\epsilon=0.0$ and $\delta=2.5^{\circ}$).\label{ta:Tamonoclinicshort}}
\begin{tabular}{c|c|c|c}
\hline
\hline
                & $\epsilon=0.0$          & $\epsilon=0.1$          & $\epsilon=0.2$ \\
                & and $\delta=0.0^{\circ}$& and $\delta=2.5^{\circ}$& and $\delta=5.0^{\circ}$\\
\hline
$\bar{\Gamma}$  & $-5.4083$, $-5.4083$    & $-5.2830$, $-5.2143$    & $-5.1671$, $-5.0339$\\
\hline
$\bar{H_2}$     & $-1.1180$, $-1.1180$    & $-1.3130$, $-0.9753$    & $-1.4975$, $-0.9464$\\
\hline
$\bar{C}$       & $-1.1180$, $-1.1180$    & $-1.3115$, $-0.9590$    & $-1.4860$, $-0.8842$\\
\hline
\hline
\end{tabular}\\
\end{table}

\begin{table}[tb]
\caption{Breakdown of the effective degeneracy of occupied states at high symmetry points on the oblique lattice for the {\em long} u.c.~slab{--Eqn.~\eqref{eq:longm}--}in eV.\label{ta:Tamonocliniclong}}
\begin{tabular}{c|c|c|c}
\hline
\hline
                & $\epsilon=0.0$          & $\epsilon=0.1$          & $\epsilon=0.2$ \\
                & and $\delta=0.0^{\circ}$& and $\delta=2.5^{\circ}$& and $\delta=5.0^{\circ}$\\
\hline
$\bar{\Gamma}$  & $-3.6056$, $-3.6056$    & $-3.5427$, $-3.4393$    & $-3.4941$, $-3.2940$\\
\hline
$\bar{H_2}$     & $-2.2361$, $-2.2361$    & $-2.3117$, $-2.1380$    & $-2.3806$, $-2.0785$\\
\hline
$\bar{C}$       & $-2.2361$, $-2.2361$    & $-2.3152$, $-2.1353$    & $-2.3913$, $-2.0717$\\
\hline
\hline
\end{tabular}\\
\end{table}

An observation from Tables \ref{ta:Tamonoclinic}, \ref{ta:Tamonoclinicshort}, and \ref{ta:Tamonocliniclong} is that {\em the degenerate states protected by $C_4$ symmetry always are located at energies away from the Fermi energy; i.e., away from zero eV}.


\begin{figure*}[tb]
\begin{center}
\includegraphics[width=0.96\textwidth]{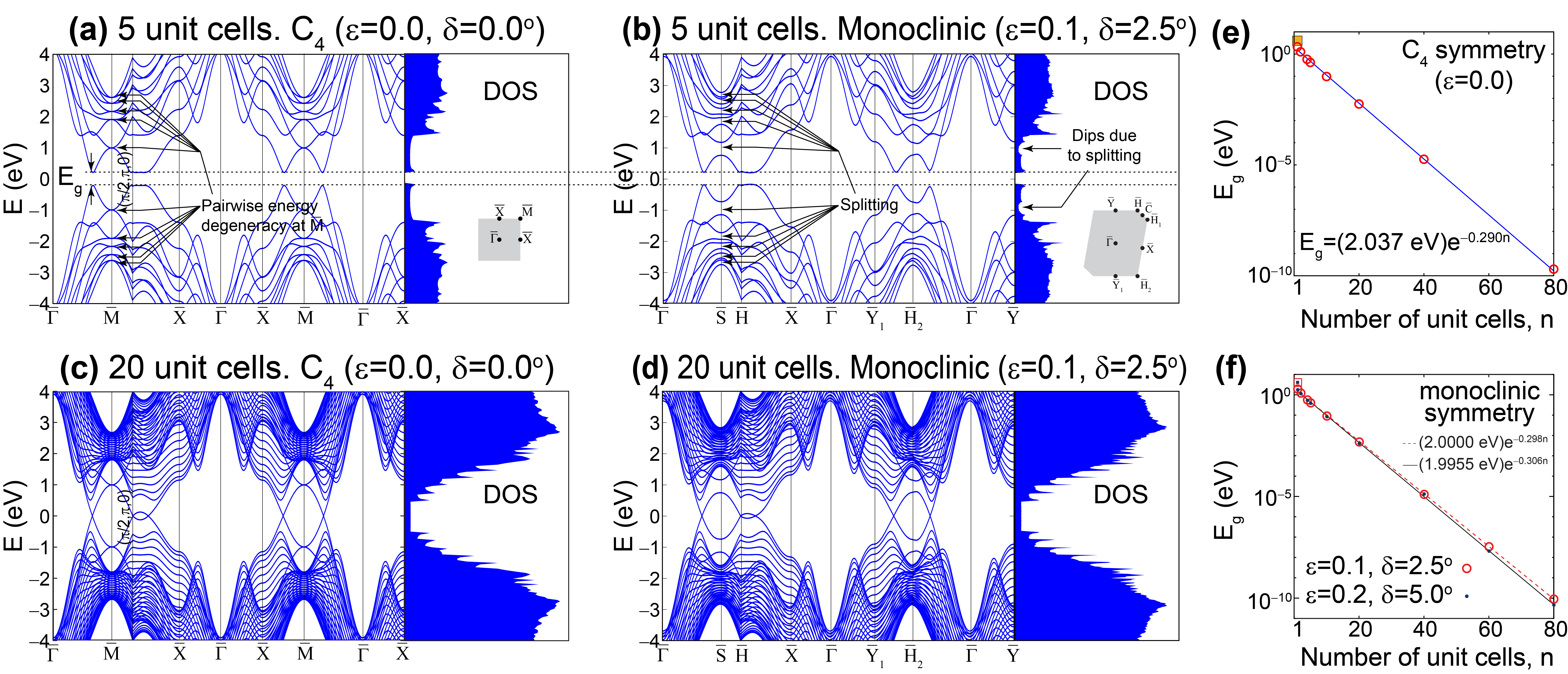}
\end{center}
\caption{(a) Electronic structure of $n=5$ u.c.~thick slabs with topological protection (TCI: $\epsilon=0$ and $C_4$ in-plane symmetry) or (b) without ($\epsilon=0.1$ and $\delta=2.5^{\circ}$ {in Eqn.~\eqref{eq:5ucm}}; no in-plane symmetry). The energy gap $E_g$ is highlighted by dashed horizontal lines, and DOS plots are added to emphasize the bandgaps. (c) and (d) Emergence of {\em zero-energy states} on a thicker slab {\em with} [subplot (c)] {\em or without topological protection} [subplot (d)]. (e) $E_g$ decays exponentially with the number $n$ of u.c.s, regardless of the presence of in-plane $C_4$ symmetry (in which topological protection exists) or the lack of it (where no topological protection exists). The single square for $n=1$ in the upper subplot is $E_g$ for the long u.c.\label{fig:newMonoclinic}}
\end{figure*}


{It is possible to build thicker oblique slabs following prescriptions similar to those leading to Eqn.~\eqref{eq:4ucs}. For example,
the oblique slab with 5 u.c.~thickness displayed on Fig.~\ref{fig:newMonoclinic}(b) is given by:
\begin{eqnarray}\label{eq:5ucm}
&H_{5uc,m}(k_x,k_y)=\\
&\left(
\begin{smallmatrix}
  H_{s,m}(k_x,k_y) & \tau & \mathcal{O} & \mathcal{O} & \mathcal{O} \\
  \tau^T & H_{s,m}(k_x,k_y) & \tau & \mathcal{O} & \mathcal{O} \\
  \mathcal{O} & \tau^T & H_{s,m}(k_x,k_y) & \tau & \mathcal{O} \\
  \mathcal{O} & \mathcal{O} & \tau^T & H_{s,m}(k_x,k_y) & \tau \\
  \mathcal{O} & \mathcal{O} & \mathcal{O} & \tau^T & H_{s,m}(k_x,k_y)
\end{smallmatrix}
\right)\nonumber
\end{eqnarray}
with $\epsilon=0.1$ and $\delta=2.5^{\circ}$.
}

Oblique slabs with finite thicknesses prove that the emergence of zero-energy states is not due to topological protection: zero-energy surface states persist even on slabs with oblique symmetry, for which the weak topological protection is long gone. Energy splitting at high-symmetry points does not arise from hybridization among $p_x$ and $p_y$ orbitals, but because block-diagonal decoupled submatrices--which are identical when $C_4$ holds--cease to be equal under a lowered symmetry.

But the lack of hybridization among $p_x$ and $p_y$ orbitals guaranteed at high symmetry $k-$points no longer holds for slab Hamiltonians at $k-$points located away from $\bar{\Gamma}$ and $\bar{M}$, giving rise to energy gaps $E_g$ highlighted by horizontal dashed lines on {Figs.~\ref{fig:newMonoclinic}(a) and \ref{fig:newMonoclinic}(b)}. In addition, $n=5$ degenerate energy crossings are highlighted on Fig.~\ref{fig:newMonoclinic}(a), and their split is seen on Fig.~\ref{fig:newMonoclinic}(b) with clarity. This is because the five occupied eigenvalues have identical values for the two independent block-diagonal submatrices for $p_x$ and $p_y$ orbitals, respectively, at high-symmetry points on the 5 u.c.-thick slab.

The bulk-boundary correspondence taking place on strong TIs guarantees the existence of zero-energy states on slabs of sufficient thickness, and one may na\"ively assume that the zero-energy states on TCI slabs--as observed on Fig.~\ref{fig:newMonoclinic}(c)--arise from topology. Nevertheless, the bulk-boundary correspondence no longer holds when the weak topology of a TCI is broken, if only because the topological protection has been removed. Yet, Fig.~\ref{fig:newMonoclinic}(d) shows beyond any possible doubt that zero-energy states remain present.

{As initially hinted at on Fig.~\ref{fig:Fi3}, the electronic band gap decays  with the number $n$ of u.c.s on TCI slabs of finite thicknesses exponentially.} Furthermore, Fig.~\ref{fig:newMonoclinic}(e) lists the electronic bandgap $E_g$ as a function of the number $n$ of u.c.s on a given slab [multiple bandgaps $E_g$ are visible on Fig.~\ref{fig:Fi3} as well]. The upper subplot on Fig.~\ref{fig:newMonoclinic}(e) summarizes the behavior of slabs with $C_4$ symmetry, while the lower subplot the behavior of slab Hamiltonians without in-plane symmetry. [The squares seen on Fig.~\ref{fig:newMonoclinic}(e) for $n=1$ correspond to the energy gap for the long u.c.] Although the slopes on the semilog plots become smaller as the elongation $\epsilon$ proceeds, the exponential closing of the zero-energy state does not distinguish among Hamiltonians with fourfold symmetry and those lacking in-plane symmetry.

The logical conclusion, thus, is that zero-energy surface states are {\em not topological} on TCIs. Zero-energy states can be induced by other means besides topology \cite{Fradkin}, and they are akin to solitons \cite{Shen}. From this point onwards, only Hamiltonians with tetragonal ($C_4$ in-plane) symmetry will be considered to prove that 1 u.c.~thick slabs can be considered TCIs on their own merit.

\section{Square-root nature of original TCI Hamiltonian and its identical topology in 3d and 2D}\label{sec:I}

The midgap states in the entanglement spectrum or in the entanglement Hamiltonian are customarily employed to distinguish between topologically trivial and non-trivial phases of TCIs \cite{PhysRevB.73.245115,PhysRevLett.104.130502,PhysRevB.87.035119,Chang_2014}. Here, on the other hand, we will rely on the square Hamiltonian of the TCI, to demonstrate an identical topological characterization at the bulk and single u.c.~limits of the TCI Hamiltonian, which serves as the first indication of the non-trivial topology of the TCI at the 2D limit.

Indeed, Arkinstall and coworkers proposed the creation of new topological phases by inducing a {\em square root} operation to topological tight-binding Hamiltonians \cite{PhysRevB.95.165109}. The square root operation provides a chiral symmetric arrangement of energy bands at positive and negative energies, inducing spectral symmetries at the expense of {\em broken crystal symmetries}. In this case, the broken symmetry is a reflection with respect to the $xy-$plane by two atoms vertically stacked within the u.c. Arkinstall {\em et al.}~also indicate that the square root Hamiltonian of a topological phase retains topological features. There is intense and ongoing work on square-root descriptions of topological phases \cite{Mizoguchi,Yoshida}.

From Eqns.~\eqref{eq:Halpha} and \eqref{eq:HAB}, $[H^A(k_x,k_y),H^{AB}(\mathbf{k})]=\mathbb{O}$ because $H^{AB}(\mathbf{k})$ [Eqn.~\eqref{eq:HAB}] is proportional to $\sigma_0$, making the square Hamiltonian of the (bulk) TCI block-diagonal:
\begin{equation}\label{eq:H2}
\left(H(\mathbf{k})\right)^2=\left( \begin{array}{cc}
H^{2}_b(\mathbf{k}) &  \mathbb{O}\\
\mathbb{O}   & H^{2}_b(\mathbf{k})\\
\end{array} \right),
\end{equation}
where $H^{2}_b(\mathbf{k})$, defined as:
\begin{eqnarray}\label{eq:H2final}
H^{2}_b(\mathbf{k})\equiv & (H^{A}(k_x,k_y))^{2}+H^{AB}(\mathbf{k})H^{AB\dagger}(\mathbf{k})\nonumber\\
= &d_{0,b}(\mathbf{k})\sigma_0+\mathbf{d}_{b}(\mathbf{k})\cdot \boldsymbol{\sigma},
\end{eqnarray}
is a renormalized Hamiltonian in which one of the two sublattices has been effectively removed \cite{Naumis_2021}, and $b$ stands for bulk. Explicit calculations yield:
\begin{equation}\label{eq:7}
d_{0,b}(\mathbf{k})=h_0(k_x,k_y)^{2}+h_x(k_x,k_y)^{2}+h_z(k_x,k_y)^{2}+|f_b(\mathbf{k})|^{2},
\end{equation}
and:
\begin{equation}\label{eq:8}
\mathbf{d}_b(\mathbf{k})=\mathbf{d}_b(k_x,k_y)=2h_0(k_x,k_y)\boldsymbol{h}(k_x,k_y),
\end{equation}
where
\begin{eqnarray}\label{eq:h0h}
&h_0(k_x,k_y)=\\
&t_1^{A} (\cos k_x +\cos k_y)+2t_2^{A} \cos k_x \cos k_y, \text{ and}\nonumber\\
&\mathbf{h}(k_x,k_y)=(2t_2^{A} \sin k_x \sin k_y,0,t_1^{A} (\cos k_x -\cos k_y)),\nonumber
\end{eqnarray}
which in turn originate from reparameterizing $H^A(k_x,k_y)$ as $h_0(k_x,k_y)\sigma_0+\mathbf{h}(k_x,k_y)\cdot \boldsymbol{\sigma}$ [and whose vector field $\mathbf{h}(k_x,k_y)$ is shown on Fig.~\ref{fig:Fi2new}(a)]. Equation \eqref{eq:8} states that $\mathbf{d}_b(\mathbf{k})=\mathbf{d}_b(k_x,k_y)$, so that {\em the vector field does not depend on $k_z$ in the bulk}, giving rise to the two-dimensional plot shown on Fig.~\ref{fig:Fi2new}(b). [Strong TIs have circulating paths enclosing the origin \cite{Asboth_2016}, which is not the case on Fig.~\ref{fig:Fi2new}.]

{  The topological nature of $H^{2}_b(\mathbf{k})$, encoded in $\mathbf{d}_b(k_x,k_y)\cdot \boldsymbol{\sigma}$ \cite{Asboth_2016}, is obtained next. For definiteness, an expression for $\mathbf{d}_b(\mathbf{k})$ for small values of $k_x$ and $k_y$ was considered; $|k_x|<< \pi/a$ and $|k_y|<<\pi/a$.  In that limit:
\begin{equation*}
h_0(k_x,k_y)\simeq 2t_1^A\left( 2-\frac{k_x^2}{2}-\frac{k_y^2}{2} +4t_2^A\left(1-\frac{k_x^2}{2}-\frac{k_y^2}{2} \right) \right),
\end{equation*}
and:
\begin{equation*}
\mathbf{h}(k_x,k_y)\simeq \left( 2t_2^Ak_xk_y,0,t_1^A\left(\frac{k_y^2}{2}-\frac{k_x^2}{2} \right) \right).
\end{equation*}

Crystal momenta $k_x$ and $k_y$ are now recast as $k_0\cos\theta$ and $k_0\sin\theta$, to determine the evolution of $H^2_b(\mathbf{k})$ under an adiabatic rotation on $k-$space by $2\pi$, as encoded on $\mathbf{d}_b(k_x,k_y)\cdot \boldsymbol{\sigma}$. To leading order, one gets around $\bar{\Gamma}$:
\begin{equation}
\mathbf{d}_b(k_0,\theta)\cdot \boldsymbol{\sigma}\to 4k_0^2\left( t_1^A+t_2^A \right)
\begin{pmatrix}
-\frac{t_1^A\cos 2\theta}{2} & t_2^A\sin 2\theta\\
t_2^A\sin 2\theta & \frac{t_1^A\cos 2\theta}{2}
\end{pmatrix},
\end{equation}
so that $\mathbf{d}_b(k_0,\theta)\cdot \boldsymbol{\sigma}$ changes by $4\pi$ when $\theta$ completes a cycle, for a winding number of 2.

\begin{figure}
\begin{center}
\includegraphics[width=0.48\textwidth]{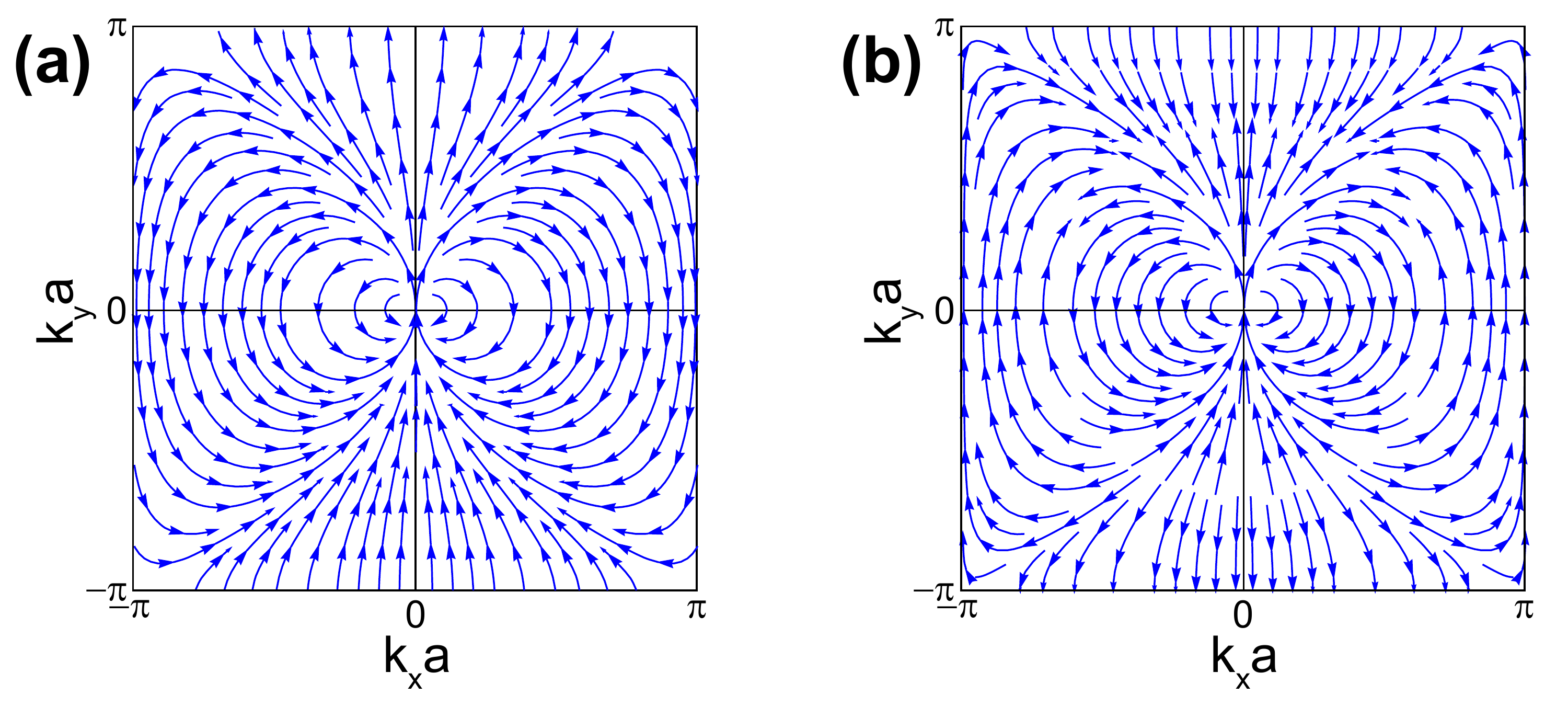}
\end{center}
\caption{(a) Vector field $\mathbf{h}(k_x,k_y)$ for $H^A(k_x,k_y)$. { (b) Vector field $\mathbf{d}_b(\mathbf{k})$ of $H^2(\mathbf{k})$; given that no terms on $H^{AB}(\mathbf{k})$ enter on it, the vector field $\mathbf{d}_b(\mathbf{k})=\mathbf{d}_b(k_x,k_y)$ stands unmodified in the bulk, or in 2D slabs with u.c.~thicknesses [$\mathbf{d}_b(k_x,k_y)=\mathbf{d}_s(k_x,k_y)=\mathbf{d}_l(k_x,k_y)$].}\label{fig:Fi2new}}
\end{figure}

The square roots of the eigenvalues from $H^{2}_b(\mathbf{k})$ yield the dispersion of the Hamiltonian [Eqn.~\eqref{eq:H2reduced}] shown in Fig.~\ref{fig:F1}(c):
\begin{equation}\label{eq:banddispersion}
E_{\pm \pm}(\mathbf{k})=\pm \sqrt{\left(h_0(k_x,k_y)\pm|\boldsymbol{h}(k_x,k_y)|\right)^{2}+|f_b(\mathbf{k})|^{2}}.
\end{equation}
The double $\pm$ symbols label each of the four bulk bands, { and the outer $\pm$ signs explicitly indicate the electron-hole symmetry of the chiral bulk Hamiltonian.}


An intriguing result from Eqns.~\eqref{eq:7}, \eqref{eq:8}, and \eqref{eq:h0h} is that the vector field $\mathbf{d}_b(k_x,k_y)$ [originally found for the {\em bulk} TCI Hamiltonian, and which turned out to be independent of $k_z$] {\em remains unchanged even when the sample is one u.c.~thick}.

The square of Eqn.~\eqref{eq:2DC4s} is:
\begin{equation}
(H_s(k_x,k_y))^2=
\begin{pmatrix}
H_s^2(k_x,k_y) & \mathbb{O}\\
\mathbb{O}     & H_s^2(k_x,k_y)
\end{pmatrix},
\end{equation}
with
\begin{eqnarray}
H_s^2(k_x,k_y)=\left[H^A(k_x,k_y)\right]^2+|f_s(k_x,k_y)|^2\sigma_0\\
=d_{0,s}(k_x,k_y)\sigma_0+\mathbf{d}_s(k_x,k_y)\cdot \boldsymbol{\sigma}.\nonumber
\end{eqnarray}
Explicit calculations yield:
\begin{eqnarray}
d_{0,s}=h_0(k_x,k_y)^2+h_x(k_x,k_y)^2+\nonumber\\
h_z(k_x,k_y)^2+|f_s(k_x,k_y)|^2,
\end{eqnarray}
with $h_0(k_x,k_y)$, $h_x(k_x,k_y)$, and $h_z(k_x,k_y)$ {\em exactly as given in Eqn.~\eqref{eq:h0h}}, and
\begin{equation}
\mathbf{d}_s(k_x,k_y)=2h_0(k_x,k_y)\mathbf{h}(k_x,k_y),
\end{equation}
where $h_0(k_x,k_y)$ and $\mathbf{h}(k_x,k_y)$, as given by Eqn.~\eqref{eq:h0h}, suffer {\em no single modification}. This way, one concludes that:
\begin{equation}
\mathbf{d}_s(k_x,k_y)=\mathbf{d}_b(k_x,k_y).
\end{equation}

Given that the topology of a given electronic phase is encoded by its vector field, the topology of the square bulk Hamiltonian must be identical to the topology of the squared 2D Hamiltonian given by Eqn.~\eqref{eq:2DC4s}, which has never been identified with TCI behavior itself as far as we know.

In fact, it does not even matter what choice of u.c.~one makes. Taking Eqn.~\eqref{eq:2DC4l} now, one sees that
\begin{equation}
(H_l(k_x,k_y))^2=
\begin{pmatrix}
H_l^2(k_x,k_y) & \mathbb{O}\\
\mathbb{O}     & H_l^2(k_x,k_y)
\end{pmatrix},
\end{equation}
with
\begin{eqnarray}
H_l^2(k_x,k_y)=\left[H^A(k_x,k_y)\right]^2+|f_l(k_x,k_y)|^2\sigma_0\\
=d_{0,l}(k_x,k_y)\sigma_0+\mathbf{d}_l(k_x,k_y)\cdot \boldsymbol{\sigma}.\nonumber
\end{eqnarray}
But
\begin{eqnarray}
d_{0,l}=h_0(k_x,k_y)^2+h_x(k_x,k_y)^2+\nonumber\\
h_z(k_x,k_y)^2+|f_l(k_x,k_y)|^2,
\end{eqnarray}
with $h_0(k_x,k_y)$, $h_x(k_x,k_y)$, and $h_z(k_x,k_y)$ {\em still as given by Eqn.~\eqref{eq:h0h}}, and
\begin{equation}
\mathbf{d}_l(k_x,k_y)=2h_0(k_x,k_y)\mathbf{h}(k_x,k_y),
\end{equation}
where $h_0(k_x,k_y)$ and $\mathbf{h}(k_x,k_y)$ yet as written down in Eqn.~\eqref{eq:h0h} still, so that
\begin{equation}
\mathbf{d}_s(k_x,k_y)=\mathbf{d}_l(k_x,k_y)=\mathbf{d}_b(k_x,k_y),
\end{equation}
and the square of the TCI phase [Eqn.~\eqref{eq:Hbulk}], and the square of two-dimensional Hamiltonians given by Eqn.~\eqref{eq:2DC4s} and Eqn.~\eqref{eq:2DC4l} encode the very same topological features, and {\em the topological characterization of the (bulk) $H^2_b(\mathbf{k})$ holds on identical at the single u.c.~thickness limit}.

We end this work by computing Pfaffians next. That last calculation will help re-emphasize the existence of a TCI phase in 2D.

\section{Topological invariants from Pfaffians of the original TCI Hamiltonian in the bulk and 2D limits}\label{sec:III}

As indicated before, midgap states in the entanglement spectrum can be employed to distinguish between topologically trivial and non-trivial phases of TCIs \cite{PhysRevB.73.245115,PhysRevLett.104.130502,PhysRevB.87.035119,Chang_2014}. On the other hand, Pfaffians are bona-fide tools for the determination of topology as well, and those will be employed in what follows.

The bulk Hamiltonian
[Eqn.~\eqref{eq:H2reduced}] takes the following generic form at high-symmetry points:
\begin{equation}\label{eq:M}
\mathcal{H}_b=
\begin{pmatrix}
\Lambda & 0 & \lambda_b & 0 \\
0 & \Lambda & 0 & \lambda_b \\
\lambda_b & 0 &-\Lambda & 0 \\
0 & \lambda_b & 0 &-\Lambda
\end{pmatrix},
\end{equation}
with real $\Lambda$ and $\lambda_b$ listed in Table \ref{ta:Ta0}. { The operator $\mathcal{H}_b$ turns out to be the Hamiltonian of two non-interacting identical dimers. To see this, we use the operator $\mathbb{P}$
\begin{equation}
    \mathbb{P}=
\begin{pmatrix}
1   & 0 & 0 & 0 \\
0 & 0   & 1 & 0 \\
0 & 1 & 0   & 0 \\
0 & 0 & 0 & 1
\end{pmatrix}
\end{equation}
to obtain:}
\begin{equation}
\mathcal{H}_b'=\mathbb{P}\mathcal{H}_b\mathbb{P}^T=
\begin{pmatrix}
\Lambda   & \lambda_b & 0 & 0 \\
\lambda_b &-\Lambda   & 0 & 0 \\
0 & 0 & \Lambda   & \lambda_b \\
0 & 0 & \lambda_b &-\Lambda
\end{pmatrix}
\end{equation}
which are the two subblocks argued for previously; the degeneracy at high-symmetry points is due to the $C_4$ in-plane symmetry of the model. The eigenvalues of $\mathcal{H}_b$ are doubly degenerate, and take on the following two values:
\begin{equation}
E_{\pm,b}=\pm \sqrt{\Lambda^2+\lambda_b^2}.
\end{equation}

\begin{table}[tb]
\caption{Parameters $\Lambda$, $\lambda_b$, and two-fold degenerate eigenvalues $E_{\pm}=\pm\sqrt{\Lambda^2+\lambda_b^2}$ at $\Gamma$, $M$, $Z$ and $A$ $k-$points for the bulk Hamiltonian. The values of $\Lambda$ and $\lambda_b$ can be directly replaced into Eqns.~\eqref{eq:V1} and \eqref{eq:V2} at these high-symmetry $k-$points. All values are given in eV.}\label{ta:Ta0}
\begin{tabular}{c|cc|cc}
\hline
\hline
                 & $\Gamma$ & $M$   & $Z$           & $A$  \\
\hline
$\Lambda$        & $3$      & $-1$  & $3$           & $-1$ \\
$\lambda_b$      & $13/2$   & $5/2$ & $5/2$         &$-3/2$\\
\hline
$E_{\pm}$        & $\pm7.1589$      & $\pm2.6926$   & $\pm3.9051$ & $\pm1.8028$\\
\hline
\hline
\end{tabular}\\
\end{table}

Slab Hamiltonians containing $n$ u.c.s can also be permuted at the $\bar{\Gamma}$ and $\bar{M}$ points by a generalization of $\mathbb{P}$, leading to $n$ two-fold degenerate crossings at these high-symmetry points. [Such phenomenology can be most clearly observed on Figs.~\ref{fig:Fi3}(c), \ref{fig:Fi3}(d), \ref{fig:Fi3}(e), \ref{fig:Fi3}(f), and \ref{fig:Fi3}(g), in which the number of bands is small.]

The two-dimensional Hamiltonian of a single $A-$sublattice slab gives the metallic eigenvalue spectrum shown in Fig.~\ref{fig:Fi3}(h), in which quadratic dispersions with energy degeneracy at the $\bar{M}$ point at $-1$ eV can still be seen with clarity. Its counterpart, the energy dispersion of the two-dimensional $B-$sublattice shown in Fig.~\ref{fig:Fi3}(i), is the negative of the dispersion for $A-$sublattice atoms, due to the relative minus sign in its hopping terms with respect to the former sublattice [see definitions for hopping terms after Eqn.~\eqref{eq:HAB}].
{  At the sublattice limit, $H^{\alpha}$ still remains diagonal at the $\bar{\Gamma}$ and $\bar{M}$ points, and it yields two-fold degenerate eigenvalues:
\begin{equation}\label{eq:MA}
\mathcal{H}^A=
\begin{pmatrix}
\Lambda & 0 \\
0 & \Lambda \\
\end{pmatrix}
\end{equation}
 at sublattice $A$, and
\begin{equation}\label{eq:MA}
\mathcal{H}^B=
\begin{pmatrix}
-\Lambda & 0 \\
0 & -\Lambda \\
\end{pmatrix},
\end{equation}
at sublattice $B$, with $\Lambda$ still given in Table \ref{ta:Ta0}. Such degeneracies, and the magnitudes of the degenerate eigenvalues are confirmed on Figs.~\ref{fig:Fi3}(h) and \ref{fig:Fi3}(i). This is why these doubly-degenerate high-symmetry points never split in energy, despite of slab thickness.

\begin{table}[tb]
\caption{Parameters $\Lambda$, $\lambda_s$ and $\lambda_l$, and their two-fold degenerate eigenvalues [$E_{\pm}=\pm\sqrt{\Lambda^2+\lambda_s^2}$ (upper row), or $E_{\pm}=\pm\sqrt{\Lambda^2+\lambda_l^2}$, (lower row)] at the $\bar{\Gamma}$, and $\bar{M}$ $k-$points for the two possible unit cell configurations. The values of $\Lambda$ and of either $\lambda_s$ or $\lambda_l$ can be directly replaced into Eqns.~\eqref{eq:V1} and \eqref{eq:V2}. All values are given in eV.\label{ta:Ta1}}
\begin{tabular}{c|cc}
\hline
\hline
                 & $\bar{\Gamma}$ & $\bar{M}$  \\
\hline
$\Lambda$        & $3$      & $-1$  \\
$\lambda_s$      & $9/2$    & $1/2$ \\
$\lambda_l$      & $2$      & $2$   \\
\hline
$E_{\pm,s}$      & $\pm5.4083$      & $\pm1.1180$\\
$E_{\pm,l}$      & $\pm3.6056$      & $\pm2.2361$\\
\hline
\hline
\end{tabular}\\
\end{table}

The {\em structure} of the bulk Hamiltonian, Eqn.~\eqref{eq:H2reduced}, at the $\Gamma$, $M$, $Z$ and $A$ $k-$points maintains its form at the $\bar{\Gamma}$ and $\bar{M}$ points in the single u.c.~limit. In the two possible 2D cases (short or long u.c.), one replaces $\lambda_b$ to either $\lambda_s$ or $\lambda_l$, whose magnitudes are listed in Table \ref{ta:Ta1}.

Setting the label $\eta$--which can either be $b$, $s$, or $l$--to describe the bulk Hamiltonian or the short or long unit cell Hamiltonians at high symmetry points, one writes down the general expression:}
\begin{equation}\label{eq:M}
\mathcal{H}_{\eta}=
\begin{pmatrix}
\Lambda & 0 & \lambda_{\eta} & 0 \\
0 & \Lambda & 0 & \lambda_{\eta} \\
\lambda_{\eta} & 0 &-\Lambda & 0 \\
0 & \lambda_{\eta} & 0 &-\Lambda
\end{pmatrix},
\end{equation}
with real $\Lambda$ and $\lambda_{\eta}$ listed in Tables \ref{ta:Ta0} and \ref{ta:Ta1}. The eigenvalues of $\mathcal{H}$ are doubly degenerate, and take on the following values:
\begin{equation}
E_{\pm,\eta}=\pm \sqrt{\Lambda^2+\lambda_{\eta}^2}.
\end{equation}
Only the two occupied (and degenerate) eigenvectors are needed to determine the topology of the Hamiltonian \cite{FUKANE11,ANDO13,BRADLYN17,CAYSSOL21}, and these have the following real entries:
\begin{equation}\label{eq:V1}
|v_{1}\rangle=\frac{1}{\sqrt{\left(\Lambda-E_{-,\eta}\right)^2+\lambda_{\eta}^2}}\left(
\begin{matrix}
-\lambda_{\eta}\\
 0\\
\Lambda-E_{-,\eta}\\
0
\end{matrix}
\right),
\end{equation}
\begin{equation}\label{eq:V2}
|v_{2}\rangle=\frac{1}{\sqrt{\left(\Lambda-E_{-,\eta}\right)^2+\lambda_{\eta}^2}}\left(
\begin{matrix}
 0\\
-\lambda_{\eta}\\
 0\\
\Lambda-E_{-,\eta}
\end{matrix}
\right),
\end{equation}
whose general structure remains invariant in the bulk and uc.~limits. This realization will simplify the calculation of Pfaffians momentarily.

\subsection{Pfaffians in 3D and 2D}
Pfaffians are next explicitly computed to provide further credence to the u.c.-thick TCI.

First:
\begin{equation}
\begin{aligned}
{\cal A}&=i \bra{u_m} \nabla_k \ket{u_m}=i \bra{\Xi u_m} \nabla_k \ket{\Xi u_m}\\
&=i \bra{u_m} \Xi^\dagger \nabla_k \Xi \ket{u_m}=i \bra{u_m} \Xi^\dagger \Xi \nabla_{-k} \ket{u_m}\\
&=i \bra{u_m} \nabla_{-k} \ket{u_m}.\\
\end{aligned}
\end{equation}
This way, ${\cal A}=-i \bra{u_m} \nabla_{k} \ket{u_m} = -{\cal A}$, so ${\cal A} = 0$ at these planes. { In the 2D limit, $k_z=0$ as well, making ${\cal A} = 0$ too.}

The reason for writing down the eigenvectors explicitly is that the sole replacement of the given magnitude of $\Lambda$, $\lambda_{\eta}$, and $E_{\pm,\eta}$ guarantees that no extra phases are inserted into these vectors at the start of the {\em open} paths [one from $\Gamma$ to $M$, and another from $Z$ to $A$ in the bulk case, and one from $\bar{\Gamma}$ to $\bar{M}$ in the single u.c.~case]. While one has freedom to choose the eigenvector's phase at one high-symmetry point in the Brillouin zone, one must follow the state continuously along the open path to another high-symmetry point in order not to add undue discontinuities on the vector's amplitudes \cite{REV3}. Furthermore, a non-arbitrary sequencing of the valence band eigenvectors at the $\Gamma$, $M$, $Z$, and $A$ ($\bar{\Gamma}$ and $\bar{M}$) points in the bulk (1 u.c.~thick slab) case becomes necessary so that no undesired permutation of degenerate states produces additional negative signs when computing the Pfaffian. To assign the proper ordering of states at high-symmetry points, the evolution of the Hamiltonian at points in $k-$space that are located infinitesimally away from the high-symmetry points was first determined. This process is explicitly delineated to break degeneracies at the $\Gamma$ and $M$ points in the bulk TCI. To leading order in $\delta$ at $k-$point $(\delta,\delta,0)$:
\begin{equation}\label{eq:deltaHGamma}
\delta \mathcal{H}_b(\Gamma)=
\delta^2\left(
\begin{matrix}
-2 &  1 & -1 & 0\\
 1 & -2 &  0 &-1\\
-1 &  0 &  2 &-1\\
 0 & -1 & -1 & 2
\end{matrix}
\right).
\end{equation}
Similarly, at $(\pi/a-\delta,\pi/a-\delta,0)$ one gets:
\begin{equation}\label{eq:delta_H_M}
\delta \mathcal{H}_b(M)=
\delta^2\left(
\begin{matrix}
 0 &  1 &  1 & 0\\
 1 &  0 &  0 & 1\\
 1 &  0 &  0 &-1\\
 0 &  1 & -1 & 0
\end{matrix}
\right).
\end{equation}

\begin{table*}[htb]
\caption{Bulk eigenvalues at locations in reciprocal space slightly away from high-symmetry points, and eigenvectors with lifted degeneracies as $\delta\to 0$. Ordered vectors (according to the lifted degeneracies) are the ones employed to compute Pfaffians with a well defined order.\label{ta:Ta2}}
\begin{tabular}{c|c||c|c}
\hline
\hline
$\Gamma+(\delta,\delta,0)$  & $M-(\delta,\delta,0)$ & $Z+(\delta,\delta,0)$ & $A-(\delta,\delta,0)$\\
\hline
$-\frac{\sqrt{205 -76\delta^2}}{2}$, $|u_1\rangle=\frac{|v_1\rangle+|v_2\rangle}{\sqrt{2}}$ &
$-\frac{\sqrt{29 + 28\delta^2}}{2}$, $|u_1\rangle=\frac{|v_1\rangle-|v_2\rangle}{\sqrt{2}}$ &
$-\frac{\sqrt{61 -44\delta^2}}{2}$,  $|u_1\rangle=\frac{ -|v_1\rangle-|v_2\rangle}{\sqrt{2}}$ &
$-\frac{\sqrt{13 - 4\delta^2}}{2}$,  $|u_1\rangle=\frac{ -|v_1\rangle+|v_2\rangle}{\sqrt{2}}$\\
$-\frac{\sqrt{205-124\delta^2}}{2}$, $|u_2\rangle=\frac{|v_1\rangle-|v_2\rangle}{\sqrt{2}}$ &
$-\frac{\sqrt{29 + 12\delta^2}}{2}$, $|u_2\rangle=\frac{|v_1\rangle+|v_2\rangle}{\sqrt{2}}$ &
$-\frac{\sqrt{61- 92\delta^2}}{2}$,  $|u_2\rangle=\frac{  |v_1\rangle-|v_2\rangle}{\sqrt{2}}$ &
$-\frac{\sqrt{13 -20\delta^2}}{2}$,  $|u_2\rangle=\frac{ -|v_1\rangle-|v_2\rangle}{\sqrt{2}}$\\
\hline
\hline
\end{tabular}\\
\end{table*}
Similar expressions can be found for $\delta \mathcal{H}_b(Z)$ and $\delta \mathcal{H}_b(A)$. We solved for the eigenvalues and eigenvectors of $\mathcal{H}_b(\Gamma)+\delta \mathcal{H}_b(\Gamma)$, $\mathcal{H}_b(M)+\delta \mathcal{H}_b(M)$, $\mathcal{H}_b(Z)+\delta \mathcal{H}_b(Z)$, and $\mathcal{H}_b(A)+\delta \mathcal{H}_b(A)$ numerically, and the {\em ordered sequence of eigenvalues and eigenvectors} is shown in Table \ref{ta:Ta2}. In addition, amplitudes of the eigenvectors that ensure the continuous evolution of the wavefunctions' amplitude across the Brillouin zone are shown in Fig.~\ref{fig:Fig2}(a) ($\Gamma-M$ path) and Fig.~\ref{fig:Fig2}(b) ($Z-M$ path).

\begin{figure}
\begin{center}
\includegraphics[width=0.48\textwidth]{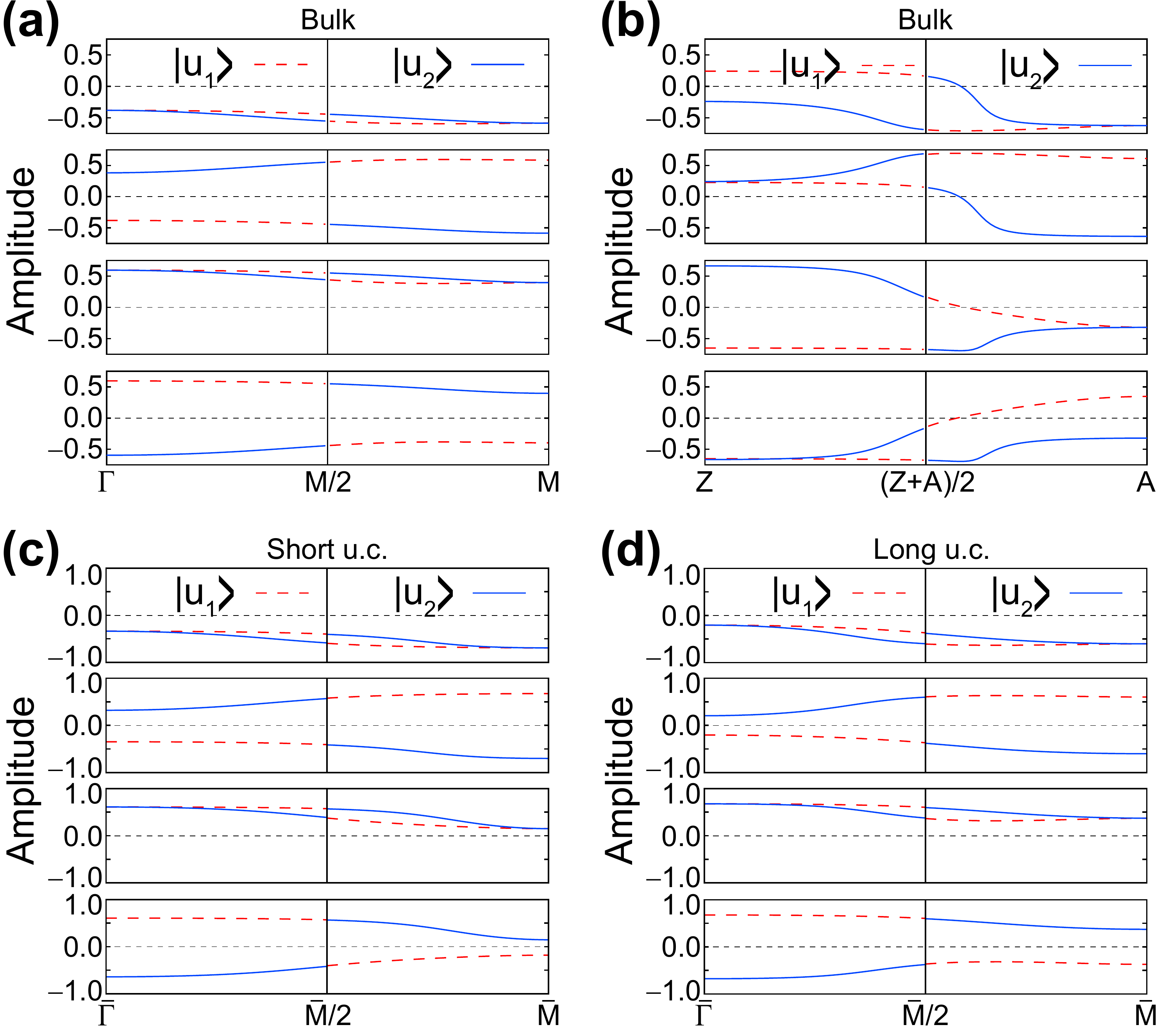}
\end{center}
\caption{Eigenvectors along the (a) $\Gamma-M$ and (b) $Z-A$ paths for the bulk TCI. (c) and (d) are eigenvectors along the $\bar{\Gamma}-\bar{M}$ paths for the short and long unit cell slabs, respectively. Despite of small differences on amplitudes, the ordering of occupied states along the bulk $\Gamma-M$ path is preserved  along the $\bar{\Gamma}$ to $\bar{M}$ path in the single u.c.~limit.\label{fig:Fig2}}
\end{figure}

{
As indicated earlier on, the twofold degeneracy persists at the 2D limit, requiring a process to order eigenstates as well. For the Hamiltonian for the short u.c.~($\eta=s$) at $k-$point $(\delta,\delta)$, $\delta H_s(\bar{\Gamma})$ is identical to the one given on Eqn.~\eqref{eq:deltaHGamma}, and similarly, $\delta H_s(\bar{M})$ is identical to Eqn.~\eqref{eq:delta_H_M} in that 2D limit.

As for the long u.c.~($\eta=l$), one gets:
\begin{equation}\label{eq:deltaHGamma}
\delta H_l(\bar{\Gamma})=
\delta^2\left(
\begin{matrix}
-2 &  1 &  0 & 0\\
 1 & -2 &  0 & 0\\
 0 &  0 &  2 &-1\\
 0 &  0 & -1 & 2
\end{matrix}
\right),
\end{equation}
and
\begin{equation}\label{eq:delta_H_M}
\delta H_l(\bar{M})=
\delta^2\left(
\begin{matrix}
 0 &  1 &  0 & 0\\
 1 &  0 &  0 & 0\\
 0 &  0 &  0 &-1\\
 0 &  0 & -1 & 0
\end{matrix}
\right),
\end{equation}
respectively.
}

{
\begin{table*}[htb]
\caption{1 u.c.~eigenvalues (in eV) at locations in reciprocal space slightly away from high-symmetry points ($\delta =0.01\frac{\pi}{a}$), and eigenvectors with lifted degeneracies as $\delta\to 0$. Vectors with lifted degeneracies are the ones employed to compute Pfaffians. Note that the ordered eigenvectors have an identical form to those listed for $k-$points $\Gamma$ and $M$ in Table \ref{ta:Ta2}.\label{ta:Ta3}}
\begin{tabular}{c|c||c|c}
\hline
\hline
$\bar{\Gamma}+(\delta,\delta)$; $\eta=s$  & $\bar{M}-(\delta,\delta)$; $\eta=s$ & $\bar{\Gamma}+(\delta,\delta)$; $\eta=l$  & $\bar{M}-(\delta,\delta)$; $\eta=l$\\
\hline
$-5.40819$, $|u_1\rangle=\frac{ |v_1\rangle+|v_2\rangle}{\sqrt{2}}$ & $-1.11817$, $|u_1\rangle=\frac{ |v_1\rangle-|v_2\rangle}{\sqrt{2}}$ &
$-3.60547$, $|u_1\rangle=\frac{ |v_1\rangle+|v_2\rangle}{\sqrt{2}}$ & $-2.23611$, $|u_1\rangle=\frac{ |v_1\rangle-|v_2\rangle}{\sqrt{2}}$\\
$-5.40808$, $|u_2\rangle=\frac{ |v_1\rangle-|v_2\rangle}{\sqrt{2}}$ & $-1.11799$, $|u_2\rangle=\frac{ |v_1\rangle+|v_2\rangle}{\sqrt{2}}$ &
$-3.60530$, $|u_2\rangle=\frac{ |v_1\rangle-|v_2\rangle}{\sqrt{2}}$ & $-2.23602$, $|u_2\rangle=\frac{ |v_1\rangle+|v_2\rangle}{\sqrt{2}}$\\
\hline
\hline
\end{tabular}\\
\end{table*}
}


We now endeavor to construct the matrix $w({\bf k})$ whose entries are given by $w_{ij}({\bf k})=\bra{u_i({\bf k})}UT\ket{u_j({\bf k})}=\bra{u_i({\bf k})}U\ket{u_j({\bf k})}$. The topological invariant is a property of the occupied bands, so $w({\bf k})$ is a $2 \times 2$ matrix. We use the states listed in Table \ref{ta:Ta2} for $|u_i({\bf k})\rangle$ for the bulk TCI. The following explicit matrices thus result,
\begin{equation}
w({\bf \Gamma})=
\begin{pmatrix}
0 & -1 \\
1 & 0
\end{pmatrix},
w({\bf M})=
\begin{pmatrix}
0 & 1 \\
-1 & 0
\end{pmatrix},
\end{equation}
and
\begin{equation}
w({\bf Z})=
\begin{pmatrix}
0 & 1 \\
-1 & 0
\end{pmatrix},
w({\bf A})=
\begin{pmatrix}
0 & 1 \\
-1 & 0
\end{pmatrix}.
\end{equation}
The Pfaffian is equal to $w_{12}({\bf k})$, yielding $(-1)^{\nu_{\Gamma M}}= \text{Pf}[w({\bf \Gamma})]/\text{Pf}[w({\bf M})]=-1$ and $(-1)^{\nu_{ZA}}=\text{Pf}[w({\bf Z})]/\text{Pf}[w({\bf A})]=1$. The topological index is given by their product:
\begin{equation}
(-1)^{\nu_{0}}=(-1)^{\nu_{\Gamma M}}(-1)^{\nu_{ZA}}=-1,
\end{equation}
and is nontrivial. { Crucially for the argument that follows, $(-1)^{\nu_{\Gamma M}}=-1$, while $(-1)^{\nu_{ZA}}=+1$.}

Table \ref{ta:Ta3} and Figs.~\ref{fig:Fig2}(c) and \ref{fig:Fig2}(d) contain the ordered eigenvalues at the $\bar{\Gamma}$ and $\bar{M}$ points for the short ($\eta=s$) and long ($\eta=l$) u.c.~thick Hamitonians; {\em the eigenvector's structure and ordering is identical to that seen on Table \ref{ta:Ta2} for the $\Gamma$ and $M$ points and on Fig.~\ref{fig:Fig2}(a)}, which leads to
$(-1)^{\nu_{\bar{\Gamma}\bar{M}}}=-1$. { $Z$ and $A$ are not defined at the 2D limit, so that
\begin{equation}\label{eq:pfaffian2d}
(-1)^{\nu_{0}}=(-1)^{\nu_{\bar{\Gamma} \bar{M}}}=-1.
\end{equation}

Along with the discussion around Fig.~\ref{fig:Fi2new} concerning the identical topology of $H^2(\mathbf{k})$ in the 3D and 2D limit, Eqn.~\eqref{eq:pfaffian2d} constitutes one of the main findings of the present work, in which we have thus proven that a single u.c.-thick slab is a TCI on its own merit.
}

\section{Conclusion}\label{sec:IV}

We considered finite-thickness slabs of a TCI Hamiltonian. We showed by explicit slab calculations--in which all in-plane symmetries were removed--that the bulk-boundary correspondence does not hold on the TCI phase. Furthermore, we relied on the chiral nature of the model to create square Hamiltonians whose topological characterization in the bulk persisted in unit cell thick slabs lacking in-plane inversion symmetry due to their bipartite nature. The identical topological characterization is further confirmed by a calculation of topological invariants using the Pfaffians at the bulk and two-dimensional limits. The approaches followed here as well as the discovery of the 2D TCI phase on the original model Hamiltonian, are novel results on this topic as far as we can tell.


\section{Acknowledgments}
Ryu Nakae initiated the research and provided motivation to complete the work contained within these pages. S.B.-L.~was funded by the U.S. Department of Energy (Award DE-SC0022120). G.N. acknowledges funding from UNAM-DGAPA (project IN102620) and CONACyT (project 1564464).  Conversations with J.W.~Villanova and K.~Chang are gratefully acknowledged.


\end{document}